\documentclass[arguments]{aastex631}
\shorttitle{V694 Peg}
\shortauthors{Hu-Shan Xu \& Li-Ying Zhu}
\graphicspath{{./}{figures/}}

\begin{document}

\title{A possible explanation of W-type phenomena in V694 Peg\footnote{Released on March, 1st, 2021}}

\author[0000-0002-7809-0347]{Hu-Shan Xu}
\affiliation{Physics Department, Yuxi Normal University, Yuxi 653100,
China}
\author[0000-0002-0796-7009]{Li-Ying Zhu}
\affiliation{Yunnan Observatories, Chinese Academy of Sciences (CAS), Kunming, P. O. Box 110, 650216, China}
\affiliation{Key Laboratory of the Structure and Evolution of Celestial Objects, Chinese Academy of Sciences, Kunming, P. O. Box 110, 650216, China}
\affiliation{Graduate University of the Chinese Academy of Sciences, Yuquan Road 19, Sijingshang Block, 100049
Beijing City, China}



\begin{abstract}

Three sets of complete multi-color light curves of V694 Peg observed in 2013, 2015 and 2019 were presented and analyzed. Our photometric solutions show that this system is an A-type shallow contact binary in 2013 and 2015, while it converted to a W-type one in 2019. A large cool spot on the component of this binary could explain the conversion, implying the W-type phenomena may be caused by magnetic activity of the components. We have collected available data of this binary and calculated 505 times of light minimum, which span 17 years. The orbital period investigation based on these timings shows there is a long-term period increase at a rate of $dP/dt$ = 4.3($\pm$ 0.3)$\times$ 10$^{-9}$ d yr$^{-1}$ superposed on a periodic variation with a period of 11.81($\pm$ 0.06) years. The cyclic orbital variation may be the result of magnetic activity cycles or the existence of a third body. Till now, only 8 transformed systems including V694 Peg have been reported. Compared with other converting contact systems between A-type and W-type, V694 Peg is recorded as the shortest-period one. All of these converting systems are late-type (later than F7) contact binaries with O'Connell effect and show cyclic period variation, which indicates that magnetic activity may be the reason for the conversion between the two types of contact binaries. For investigating the nature of A-type and W-type phenomena, the discovery of more converting contact binaries is essential.

\end{abstract}



\section{Introduction} \label{sec:intro}
Contact binaries are close binary systems with two components filling their critical Roche lobes and sharing a common envelope. They usually display the EW-type light curves. They follow a famous period-color relationship \citep[e.g.][]{Eggen1967,Qian2020}. The shorter the orbital period is, the lower the effective temperature of the binary star, thus belonging to late-type stars with strong magnetic activities due to their deeper convection zone and fast rotational velocity. \cite{Binnendijk1970} divided contact binaries into two sub-types (i.e. A-type and W-type) by the difference in the light curve. The more-massive components with higher effective temperature are A-type stars; the less-massive components with higher effective temperature are W-type.
Some researchers have investigated whether there is an evolutionary relationship between the two sub-types.

Statistical studies show that the A-type system has lower density, greater total mass and/or angular momentum than the W-type system (\citealt{Mochnacki1981,Maceroni1996,Gazeas2006}). The W-type system is characterized by late spectral type and shallow contact (\citealt{Mochnacki1973,Rucinski1974}). To explain the different characteristics of these two sub-types, researchers put forward different views: 1. The thickness of common envelope is considered to cause the difference (\citealt{Lucy1973,Mochnacki1973,Rucinski1974,Wilson1978}); 2. Different mass/energy transfer rates will lead to these two different evolutionary results (\citealt{Gazeas2006,Gazeas2008}); 3. The contact binary system has undergone the mass ratio reversal. The system with an initial mass greater than 1.8$M_\odot$ evolves into A-type, while the system with an initial mass lower than this mass evolves into W-type (\citealt{Yildiz2013}). 

The evolutionary relationship between these two sub-types is still uncertain. Furthermore, there is an obvious overlap between the distribution of the two sub-types. Importantly, whether or not there is an evolutionary link between the A- and W-type, it is difficult to observe the conversion between them. However, A few contact binaries have been detected converting between A- and W-type. Some even converse from one sub-type to another and then return to the original sub-type, such as RZ Com \citep[e.g.][]{1983Ap&SS..92...99J,1983MNRAS.203....1M,2004NewA....9..273X,2005PASJ...57..977Q,2008ChJAA...8..465H} and FG Hya \citep[e.g.][]{1972MNRAS.156..243M,1979MNRAS.189..907T,yang2000,Qian+Yang2005}. Another explanation is that to assume the presence of dark spots on the more-massive component, causing the average surface temperature to drop, thereby forming the W-type \citep[][]{Binnendijk1970, Mullan1975, Eaton1980, Stepien1980}. These systems have indeed been observed with distinct spot activity. However, dark spot activity is very common and active in late-type contact binaries, which hardly explains why observed conversion systems between the A- and W-type are so rare. So far, there are no models or direct observational evidence to support this assumption. In the present paper, a new such system V694 Peg is presented.

V694 Peg was detected by SuperWASP as an eclipsing binary system with the difference between the two minimum depths of 0.13 mag \citep[][]{Norton2011}. Thus its light curve is more like an EB-type light variation. The \emph{V} magnitude of V694 Peg is roughly 14.49 mag and the period is 0.22484 days. \citet{Lohr2012} revised its period to be 19426.310 seconds by using a sinusoidal function to fit the light curves, and found that the orbital period of V694 Peg did not change significantly in seven years. The first multi-color photoelectric observation light curves and light ephemeris MinI HJD = 2,456,533.4807(5) + 0$^d$.2248416 * \emph{E} was given by \citet{2016AJ....152...57D}. He concluded that V694 Peg belongs to a W-type contact binary with large dark spots covering. His light curves show EW-type light variation with similar depths of the two minimum. In this paper, we presented a new CCD multi-color photometric light curves. Together with previous light curves and available times of light minimum, the period variation, the changes of the light curves, the properties of the converting between the A- and W-type, the structure and evolution of V694 Peg are discussed. \\

\section{PHOTOMETRIC OBSERVATIONS AND ANALYZES}\label{sec:style}
\subsection{Observation}
\label{sec:maths} 
V694 Peg was observed on 8, 9, and 17 June 2019 with the ANDOR 2048 $\times$ 2048 CCD photometric system (the field view of 27\arcmin $\times$ 32\arcmin.) attached to the 85cm telescope (Cassegrain reflecting telescope) at the Xinglong Station of the National Astronomical Observatories. The classic broadband Johnson Cousins filters were used. The observed images were reduced by using IRAF. Two stars that are very close to the target were selected as comparison star and check star, with coordinates $\alpha_{2000}$  = 21$^{h}$28$^{m}$18.86$^{s}$, $\delta_{2000} $ = +15$^{\circ}$13$\arcmin$22.3$\arcsec$  and $\alpha_{2000}$  = 21$^{h}$28$^{m}$27.64$^{s}$, $\delta_{2000} $ = +15$^{\circ}$12$\arcmin$19.1$\arcsec$, respectively. 
The standard deviations of the differential magnitude between the comparison star and the check star are 0.018 mag  for \emph{B} band, 0.010 mag for \emph{V} band, 0.013 mag for \emph{R}$_{c}$ band and 0.013 mag for \emph{I}$_{c}$ band, which indicate that our measurements have relatively higher precision with a mean error of 0.014 mag. Our CCD Observations of V694 Peg are available in online supplementary material. Based on these data, new times of light minimum was obtained and listed in Table \ref{tab:2}.
\begin{deluxetable*}{ccc}
\tablecaption{New Times of Minimum Light.\label{tab:2}}
\tablewidth{0pt}
\tablehead{
Time of minimum [HJD](Errors) & Filters & Type}
\decimalcolnumbers
\startdata
      2458644.29796( .00014) & \emph{B} & I \\
      2458643.28513( .00023) & \emph{B} & II \\
      2458652.27920( .00052) & \emph{B} & II \\
      2458644.29794( .00010) & \emph{V} & I \\
      2458643.28538( .00021) & \emph{V} & II \\
      2458652.27855( .00026) & \emph{V} & II \\
      2458644.29776( .00014) & \emph{R}$_c$ & I \\
      2458643.28558( .00018) & \emph{R}$_c$ & II \\
      2458652.28054( .00020) & \emph{R}$_c$ & II \\
      2458644.29772( .00014) & \emph{I}$_c$ & I \\
      2458643.28558( .00019) & \emph{I}$_c$ & II \\
      2458652.27914( .00027) & \emph{I}$_c$ & II \\
\enddata
\tablecomments{I and II refers to the primary minimum and secondary minimum, respectively. }
\end{deluxetable*}

\begin{figure}
 \begin{center}
  \includegraphics[width=12cm]{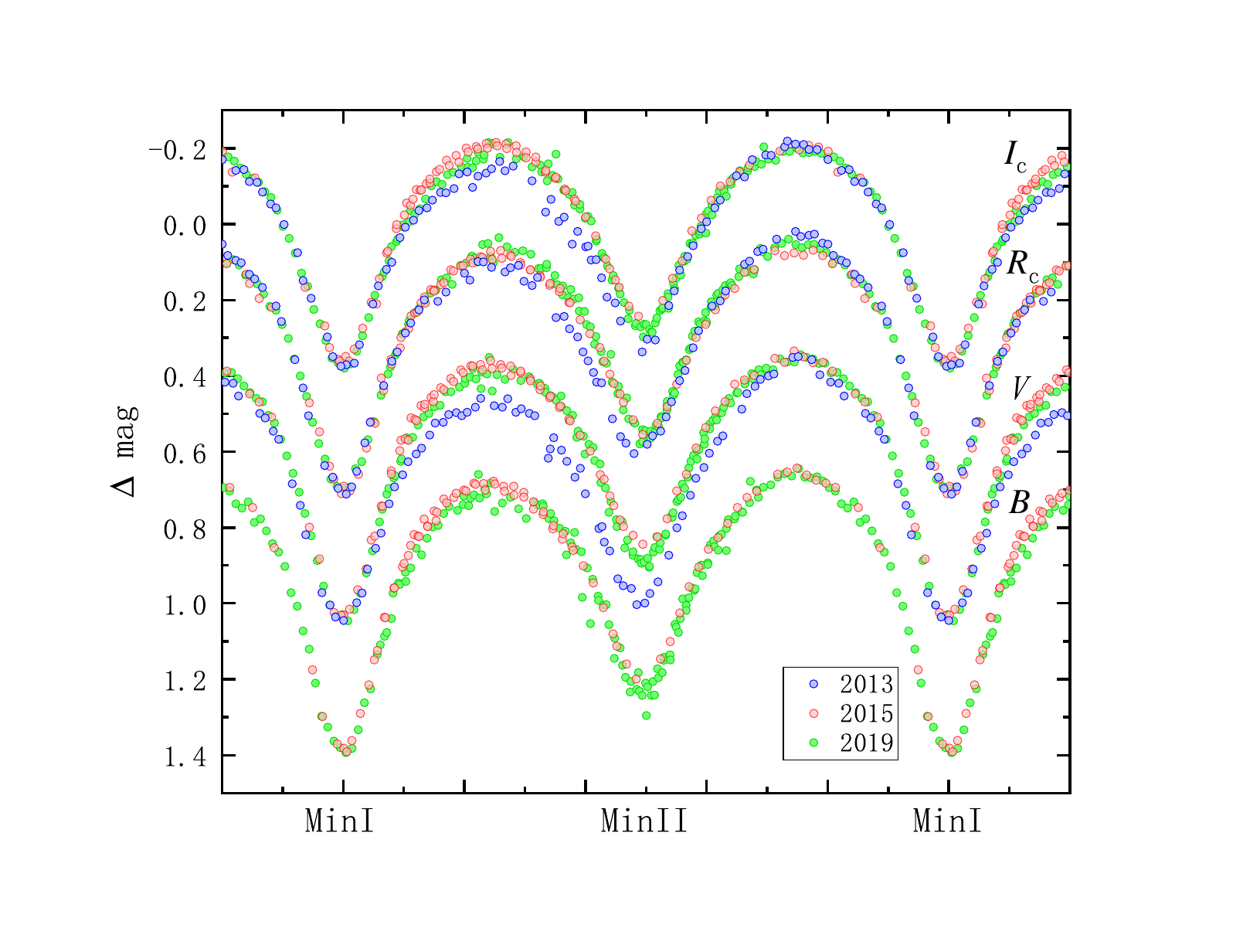}
 \end{center}
\caption{The three-sets light curves of the V694 Peg observed in 2013 (blue dots), 2015 (red dots) and 2019 (green dots).}\label{fig:3}
\end{figure}

\begin{figure}
 \begin{center}
  \includegraphics[width=11cm]{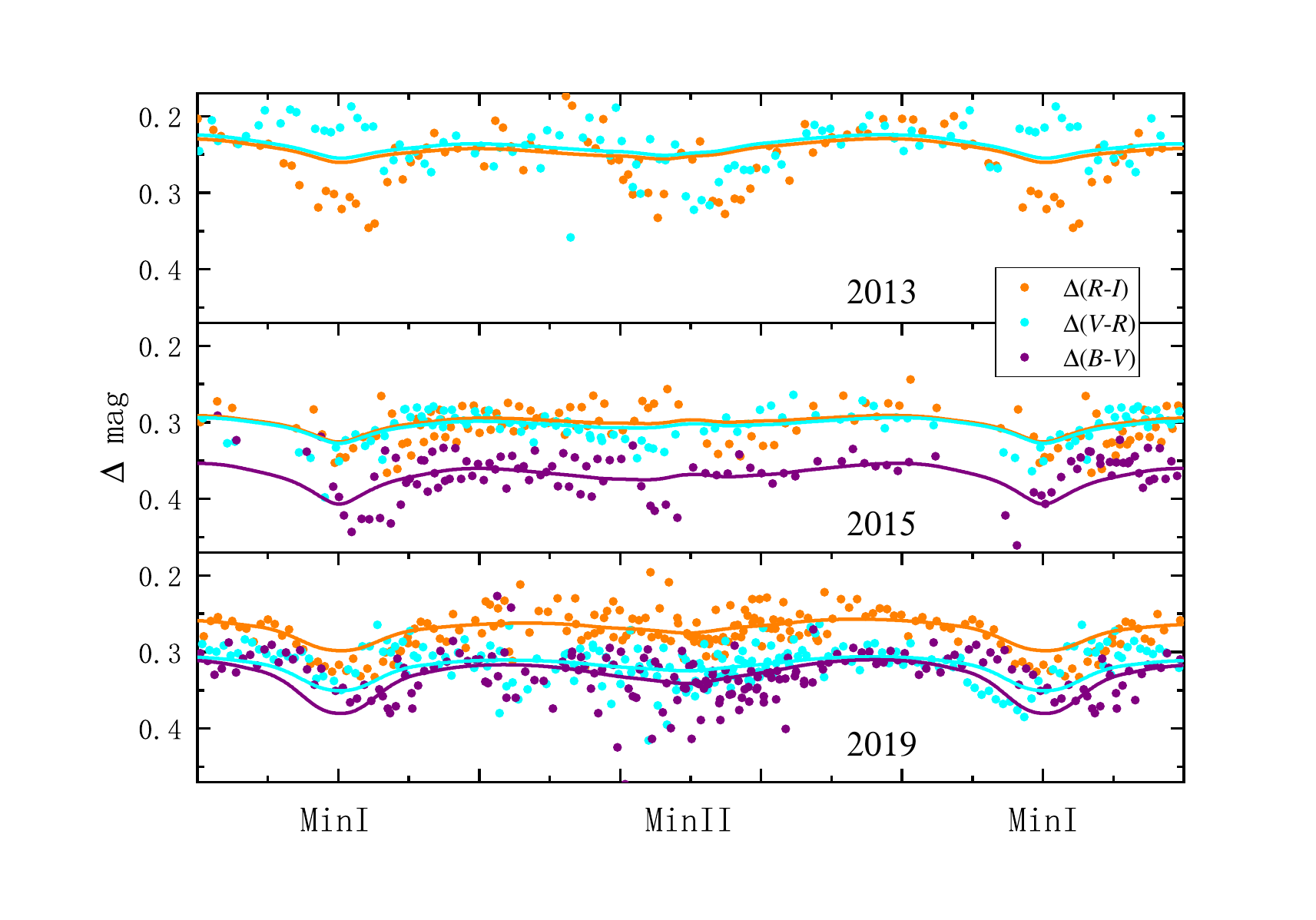}
 \end{center}
\caption{The color index curves of the V694 Peg in 2013, 2015 and 2019. The solid points represent the observed color index, and the solid lines represent the theoretical color index curves.}\label{CC}
\end{figure}

In addition, V694 Peg has also been observed by other researchers in 2013 and 2015. The 2013 light curves was observed by Djura\v{s}evi\'{c} with the 1.88 m telescope of the Kottamia Observatory in Egypt. The light curves of 2015 were observed by Chris Koen with STE4 CCD camera mounted on the 1.0 m telescope of the South African Astronomical Observatory (SAAO). Combined these light curves with ours, we have three-sets multi-color light curves which were observed in 2013, 2015 and 2019. All of them  were shown in Figure \ref{fig:3}. From Figure \ref{fig:3}, we can see that the light curves have changed from year to year. In order to see the changes clearly, we calculated the different light levels of these light curves and listed them in Table \ref{JDJX}. From the Table \ref{JDJX}, it can be noted that the difference of eclipse depth between the primary minimum and the secondary minimum in 2013 is significantly smaller than 2015 and 2019. The average magnitude of the different minimum depth is 0.06 mag in 2013, 0.17 mag in 2015 and 0.14 mag in 2019, respectively. So the light curves are more like an EW-type in 2013, and deviates from EW-type in 2015 and 2019. There is an obvious O'Connell effect \citep[][]{Connell1951} with the primary maximum lower than the secondary ones in all years. The differential magnitude between the two maximum is getting larger for shorter wavelength. It can also be seen from Figure \ref{fig:3} that the variable O'Connell effect are seen only through MaxI and fading out after MinII. This could mean that spot activity is concentrated in this region. We have also listed the difference between the maximum and minimum of color index in Table \ref{JDJX}. The color index can reflect the temperature difference between primary and secondary star to some extent. However, due to the different equipment used and the limitation of data accuracy, we only use the difference of color index for a simple discussion. In different years, the difference of color index between the primary minimum and the secondary minimum is slightly different, but there is no obvious change (such as from positive to negative), and the primary maximum and the secondary maximum are basically the same. Overall, the color index for 2015 and 2019 are relatively consistent, with some bands slightly different from them in 2013, such as the $\Delta(V-R)$ for MinI-MinII. 
In order to more intuitively compare the change of color index, we also draw the color index curves during the orbit in Figure \ref{CC}. As can be seen from Figure \ref{CC}, the overall change of $\Delta(V-R)$ in 2019 is more obvious, almost overlapping with $\Delta(B-V)$. The color index in 2013 seems to fluctuate more in the minimum and maximum parts. To avoid luminosity skew due to different selection of comparison stars, we calculate the color index $\Delta(B-V)$ for 2013 and 2019 comparison stars using data from the UCAC 4. They are 0.832 and 0.801, respectively, which are close to the color index $\Delta(B-V)= 1.157$ of the target star, and all are K-type stars. Multiple comparison stars were selected in 2015, making it difficult to accurately calculate the color index. Based on its color index curves, we assume that the situation should be similar.

\begin{deluxetable*}{cccccc}
\tablecaption{The different light levels and color index of V694 Peg. \label{JDJX}}
\tablewidth{0pt}
\tablehead{
\colhead{Year} & \colhead{Filters}& \colhead{MinI-MinII [mag]} & \colhead{MaxI-MaxII [mag]}& \colhead{MaxI-MinI [mag]}& \colhead{MaxII-MinII [mag]}}
\decimalcolnumbers
\startdata
	&	$	\Delta\emph{V}\nonumber	$	&	0.049(13)	&	0.099(12)	&	-0.567(13)	&	-0.618(12)	\\
	&	$	\Delta\emph{R}_{c}\nonumber	$	&	0.096(17)	&	0.083(08)	&	-0.576(17)	&	-0.563(09)	\\
	&	$	\Delta\emph{I}_{c}\nonumber	$	&	0.051(19)	&	0.065(09)	&	-0.510(14)	&	-0.52(13)	\\
2013	&	$	\Delta\emph{(B-V)}_{KO}\nonumber	$	&	-	&	-	&	-	&	-	\\
	&	$	\Delta\emph{(V-R)}_{KO}\nonumber	$	&	-0.047(31)	&	0.016(21)	&	0.009(30)	&	-0.055(21)	\\
	&	$	\Delta\emph{(R-I)}_{KO}\nonumber	$	&	0.046(36)	&	0.019(18)	&	-0.067(32)	&	-0.040(22)	\\
	&	$	\Delta\emph{(B-I)}_{KO}\nonumber	$	&	-	&	-	&	-	&	-	\\
	&	$	\Delta\emph{(V-I)}_{KO}\nonumber	$	&	-0.002(32)	&	0.035(21)	&	-0.058(27)	&	-0.094(26)	\\
    \hline 
	&	$	\Delta\emph{B}\nonumber	$	&	0.194(21)	&	0.038(05)   	&	-0.697(18)	&	-0.541(08)	\\
	&	$	\Delta\emph{V}\nonumber	$	&	0.197(31)	&	0.028(05)   	&	-0.660(30)	&	-0.491(08)	\\
	&	$	\Delta\emph{R}_{c}\nonumber	$	&	0.143(23)	&	0.017(05)   	&	-0.601(20)	&	-0.474(09)	\\
	&	$	\Delta\emph{I}_{c}\nonumber	$	&	0.134(25)	&	0.003(07)	&	-0.570(22)	&	-0.439(11)	\\
2015	&	$	\Delta\emph{(B-V)}_{SAAO}\nonumber	$	&	-0.003(53)	&	0.010(11) 	&	-0.037(48)	&	-0.050(16)	\\
	&	$	\Delta\emph{(V-R)}_{SAAO}\nonumber	$	&	0.054(55)	&	0.011(11) 	&	-0.059(50)	&	-0.016(17)	\\
	&	$	\Delta\emph{(R-I)}_{SAAO}\nonumber	$	&	0.009(49)	&	0.014(13) 	&	-0.030(42)	&	-0.036(20)	\\
	&	$	\Delta\emph{(B-I)}_{SAAO}\nonumber	$	&	0.060(47)	&	0.035(12) 	&	-0.127(41)	&	-0.102(19)	\\
	&	$	\Delta\emph{(V-I)}_{SAAO}\nonumber	$	&	0.063(58)	&	0.025(13) 	&	-0.090(52)	&	-0.052(18)	\\
     \hline
	&	$	\Delta\emph{B}\nonumber	$	&	0.156(14)	&	0.048(11)	&	-0.678(13)	&	-0.570(11)	\\
	&	$	\Delta\emph{V}\nonumber	$	&	0.157(09)	&	0.038(06)	&	-0.645(10)	&	-0.526(06)	\\
	&	$	\Delta\emph{R}_{c}\nonumber	$	&	0.146(11)	&	0.014(07)	&	-0.618(10)	&	-0.486(08)	\\
	&	$	\Delta\emph{I}_{c}\nonumber	$	&	0.101(10)	&	0.007(07)	&	-0.559(10)	&	-0.464(07)	\\
2019	&	$	\Delta\emph{(B-V)}_{NAOC}\nonumber	$	&	-0.001(23)	&	0.011(17) 	&	-0.033(23)	&	-0.044(17)	\\
	&	$	\Delta\emph{(V-R)}_{NAOC}\nonumber	$	&	0.010(21)	&	0.023(14) 	&	-0.027(20)	&	-0.040(15)	\\
	&	$	\Delta\emph{(R-I)}_{NAOC}\nonumber	$	&	0.045(21)	&	0.007(14) 	&	-0.059(20)	&	-0.021(15)	\\
	&	$	\Delta\emph{(B-I)}_{NAOC}\nonumber	$	&	0.054(23)	&	0.041(18) 	&	-0.119(23)	&	-0.106(18)	\\
	&	$	\Delta\emph{(V-I)}_{NAOC}\nonumber	$	&	0.055(19)	&	0.031(13) 	&	-0.086(20)	&	-0.062(13)	\\
\enddata
\end{deluxetable*}

\subsection{Photometric analyzes}
\subsubsection{Period investigations}
\begin{figure}
 \begin{center}
  \includegraphics[width=8cm]{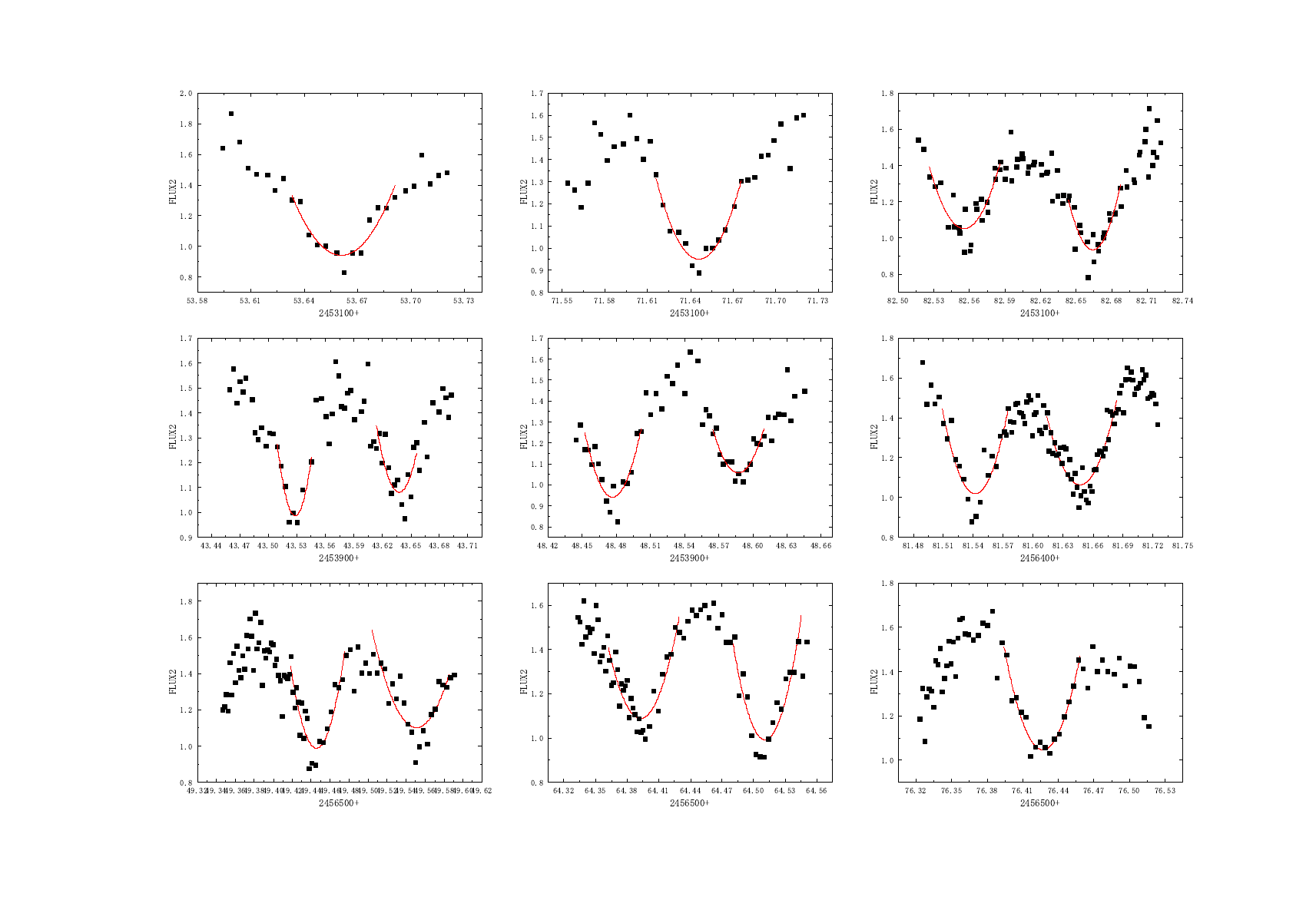}
 \end{center}
\caption{The fitting examples of the SuperWASP eclipse profiles.}\label{fig:ep}
\end{figure}

 In order to investigate the period changes of V694 Peg, we need eclipse timings spanning as longer as possible. Beside the three new times of light minimum obtained by our observations, we have derived four minimum times by fitting the eclipse profiles of 2013 and 2015 observations. In addition, the data observed by the SuperWASP Survey covering from 2004 to 2014 are very useful and kindly provided by Prof. Norton. Based on these data, we have found 482 eclipse profiles and calculated the corresponding timings by parabolic fitting. Some examples of these fittings were shown in Figure \ref{fig:ep}. Most timings derived from SuperWASP data have the errors smaller than 0.003 days. For other sky survey missions with low time resolution, such as the Catalina Sky Survey (CSS, \citealt{2014ApJS..213....9D}), the All-Sky Automated Survey for SuperNova (ASAS-SN, \citealt{2014ApJ...788...48S,2017PASP..129j4502K}) and the Zwicky Transient Facility \citep[ZTF,][]{2019PASP..131g8001G,2019PASP..131a8003M}, we used the mean minimum time method (\citealt{2021ApJ...922..122L}) to obtain the complete light curve. In all, we have a total of 505 available minimum times. All of them are listed in Table \ref{tab:2} with the order from near to far. Because it is a long table, only the first 40 timings are listed in the text. The epoch numbers and O-C values computed with the following ephemeris are listed in the third and forth columns of Table \ref{tab:4}, respectively.
 \begin{equation}\label{111}
   Min.I(HJD)= 2458644.29785 + 0^d.2248416 * E.
 \end{equation}

\begin{deluxetable*}{ccccc}
\tablecaption{All available times of light minimum of V694 Peg.\label{tab:4}}
\tablewidth{0pt}
\tablehead{ HJD & Errors & Epoch number & O-C & source }
\decimalcolnumbers
\startdata
		2458643.28542 	&	0.00020 	&	-4.5	&	-0.000640525	&	NAOC 85cm	\\
		2458644.29785 	&	0.00013 	&	0	&	0	&	NAOC 85cm	\\
		2458652.27936 	&	0.00031 	&	35.5	&	-0.000362525	&	NAOC 85cm	\\
		2457272.42494 	&	0.00028 	&	-6101.5	&	-0.002187675	&	SAAO 1.0 m	\\
		2457281.30711 	&	0.00022 	&	-6062	&	-0.0012589	&	SAAO 1.0 m	\\
		2456533.47985 	&	0.00017 	&	-9388	&	-0.005526934	&	\cite{2016AJ....152...57D}	\\
		2456533.36517 	&	0.00022 	&	-9388.5	&	-0.007786158	&	\cite{2016AJ....152...57D}	\\
		2453891.14143 	&	0.00065 	&	-21140	&	-0.006052623	&	CSS	\\
		2453891.25487 	&	0.00043 	&	-21139.5	&	-0.005030792	&	CSS	\\
		2455021.64492 	&	0.00048 	&	-16112	&	-0.005870974	&	CSS	\\
		2455021.75784 	&	0.00069 	&	-16111.5	&	-0.005372359	&	CSS	\\
		2456195.31303 	&	0.00074 	&	-10892	&	-0.010649075	&	CSS	\\
		2457049.49145 	&	0.00067 	&	-7093	&	-0.00528085	&	ASAS-SN	\\
		2457049.60392 	&	0.00059 	&	-7092.5	&	-0.005231625	&	ASAS-SN	\\
		2457768.65106 	&	0.00055 	&	-3894.5	&	-0.001368525	&	ASAS-SN	\\
		2458344.36043 	&	0.00049 	&	-1334	&	0.0012127	&	ASAS-SN	\\
		2458362.90805 	&	0.00014 	&	-1251.5	&	-0.000595175	&	ZTF	\\
		2458363.02068 	&	0.00014 	&	-1251	&	-0.00038595	&	ZTF	\\
		2458728.05018 	&	0.00027 	&	372.5	&	-0.001142375	&	ZTF	\\
		2459091.50729 	&	0.00016 	&	1989	&	-0.00039795	&	ZTF	\\
		2459091.62004 	&	0.00032 	&	1989.5	&	-6.8725E-05	&	ZTF	\\
		2459349.96067 	&	0.00043 	&	3138.5	&	-0.002379675	&	ZTF	\\
		2459350.07319 	&	0.00018 	&	3139	&	-0.00228045	&	ZTF	\\
		2453137.70115 	&	0.00056 	&	-24491	&	-0.00229395	&	SuperWASP	\\
		2453138.70572 	&	0.00098 	&	-24486.5	&	-0.009510925	&	SuperWASP	\\
		2453151.63774 	&	0.00121 	&	-24429	&	-0.00588005	&	SuperWASP	\\
		2453152.65016 	&	0.00111 	&	-24424.5	&	-0.005247025	&	SuperWASP	\\
		2453153.66094 	&	0.00111 	&	-24420	&	-0.006254	&	SuperWASP	\\
		2453153.65985 	&	0.00082 	&	-24420	&	-0.007344	&	SuperWASP	\\
		2453155.68360 	&	0.00147 	&	-24411	&	-0.00716795	&	SuperWASP	\\
		2453159.61228 	&	0.00387 	&	-24393.5	&	-0.013215075	&	SuperWASP	\\
		2453162.65710 	&	0.00083 	&	-24380	&	-0.003756	&	SuperWASP	\\
		2453164.67706 	&	0.00362 	&	-24371	&	-0.00736995	&	SuperWASP	\\
		2453165.57784 	&	0.00150 	&	-24367	&	-0.00595615	&	SuperWASP	\\
		2453165.69182 	&	0.00217 	&	-24366.5	&	-0.004396925	&	SuperWASP	\\
		2453166.58630 	&	0.00175 	&	-24362.5	&	-0.009283125	&	SuperWASP	\\
		2453168.61196 	&	0.00212 	&	-24353.5	&	-0.007197075	&	SuperWASP	\\
		2453169.62457 	&	0.00082 	&	-24349	&	-0.00637405	&	SuperWASP	\\
		2453170.63433 	&	0.00082 	&	-24344.5	&	-0.008401025	&	SuperWASP	\\
		2453171.64602 	&	0.00067 	&	-24340	&	-0.008498	&	SuperWASP	\\
	...             & ...           & ...       & ...  &  ... \\
\enddata
\end{deluxetable*}

\begin{figure}
 \begin{center}
  \includegraphics[width=12cm]{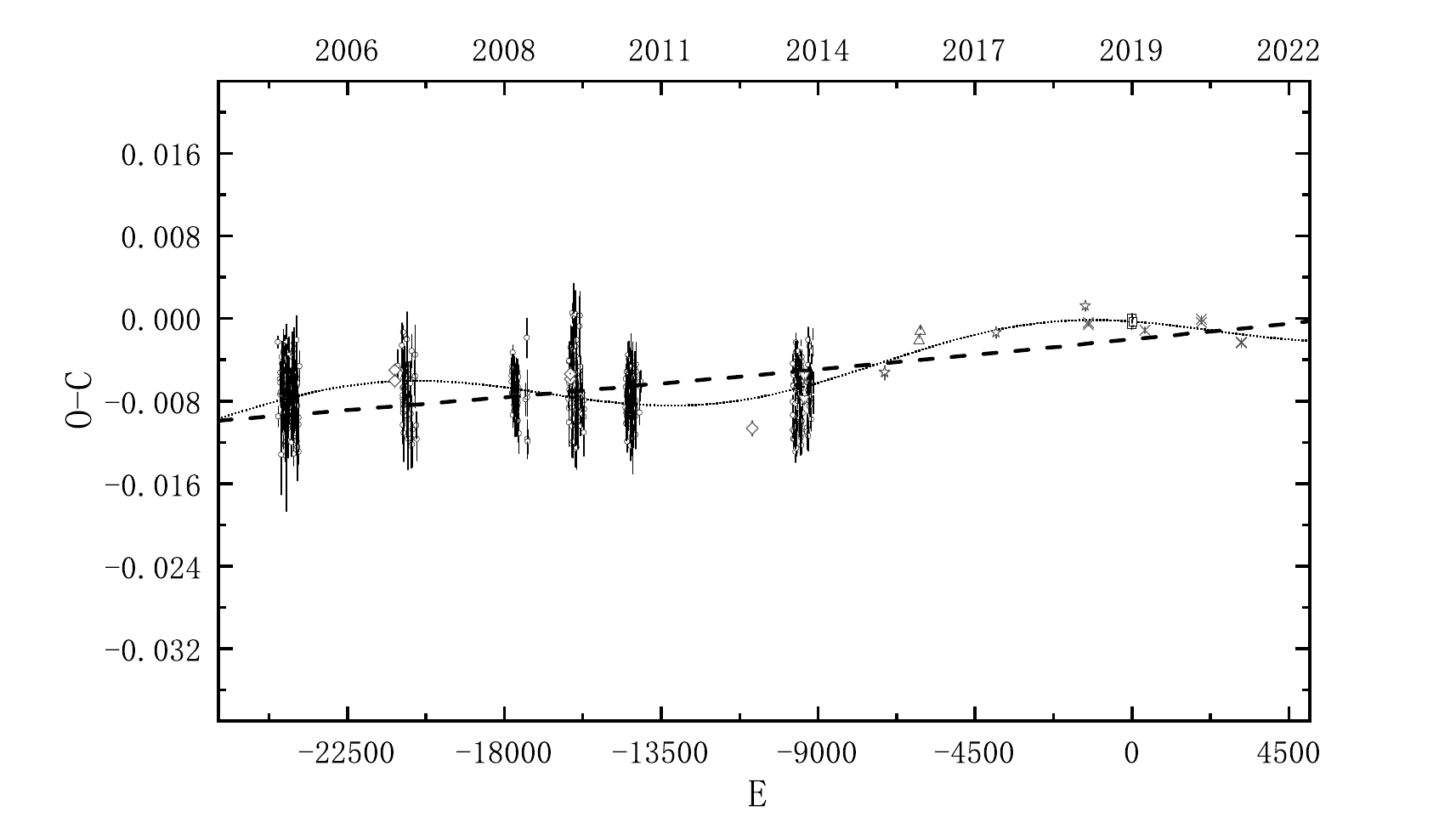}
 \end{center}
\caption{O-C diagram of the V694 Peg. The squares, triangles, inverted triangles, circles, diamonds, pentagrams, and crosses are minimum times from NAOs 85cm, KOs 1.88m, SAAO 1.0m, SuperWASP, CSS, ASAS-SN, and ZTF respectively. The dashed lines are the quadratic fit, and the solid lines are combination of the quadratic fit and sinusoidal fit. }\label{OC}
\end{figure}

The O-C diagram is shown in the Figure \ref{OC}, in which the squares, triangles, inverted triangles, circles, diamonds, pentagrams, and crosses represent minimum times from NAOs 85cm, KOs 1.88m, SAAO 1.0m, SuperWASP, CSS, ASAS-AN, and ZTF respectively. Weights of 1/$\sigma^2$ were assigned to the data, where $\sigma$ is the error of the times of light minimum. By means of a least-squares method, The following ephemeris was obtained, which means the O-C curve of V694 Peg could be fitted well by combining the quadratic term and the sinusoidal term, see the solid line of the Figure \ref{OC}.
\begin{eqnarray}\label{P}
   & Min I  =  HJD\ 2458644.29582(1) + 0^d.224841855(2)*E \nonumber \\
   & + 0.13(1)*10^{-11}* E^{2} \nonumber \\
   & + 0.002432(1)* sin(0.01877(9)\deg * E +135.6(2)\deg)
\end{eqnarray}

The quadratic term implies that the period of V694 Peg is increasing at a rate of 4.3(3)$\times$10$^{-9}$d yr$^{-1}$. The corresponding fit curve are shown in the Figure \ref{OC} with the dashed line. The sinusoidal term indicates that the orbital period of the system oscillates periodically with an amplitude of 0.002432(1) days and a period of 11.81(6) years.

\subsubsection{Light curve analyzes}

We used Wilson-Devininney (W-D) method \citep[][]{Wilson+Devinney1971, Wilson1990, Wilson1994} to analyze the three-sets of light curves and obtain the photometric parameters in different years. The effective temperature of the star 1 was fixed as 4450 K, which was derived by \citet{Koen2016}). The gravity-darkening coefficients were assumed as $g_{1} = g_{2} = 0.32$ \citep[][]{Lucy1976} and the bolometric albedo as $A_{1} = A_{2} = 0.5$ \citep[][]{Rucinski1969} due to the late-type components of V694 Peg. Bolometric and bandpass square-root limb darkening parameters were taken from \citet{VanHamme1993} : $x_{1bol}=0.313, y_{1bol}=0.366,x_{1B}=1.176	,y_{1B}=-0.388,x_{1V}=0.785,y_{1V}=0.016,x_{1Rc}=0.545	,y_{1Rc}=0.240,x_{1Ic}=0.356,y_{1Ic}=0.360
,$ the corresponding parameters of the star 2 are selected with different values according to different temperatures.  The adjustable parameters are the inclination $i$, the effective temperature $T_{2}$ of the star 2, the monochromatic luminosity $L_{1B},L_{1V},L_{1Rc}$ and $L_{1Ic}$ of the star 1, and the dimensionless potentials $\Omega$$_{1}$ and $\Omega$$_{2}$ of each component.

\begin{figure*}
 \begin{center}
  \includegraphics[width=18cm]{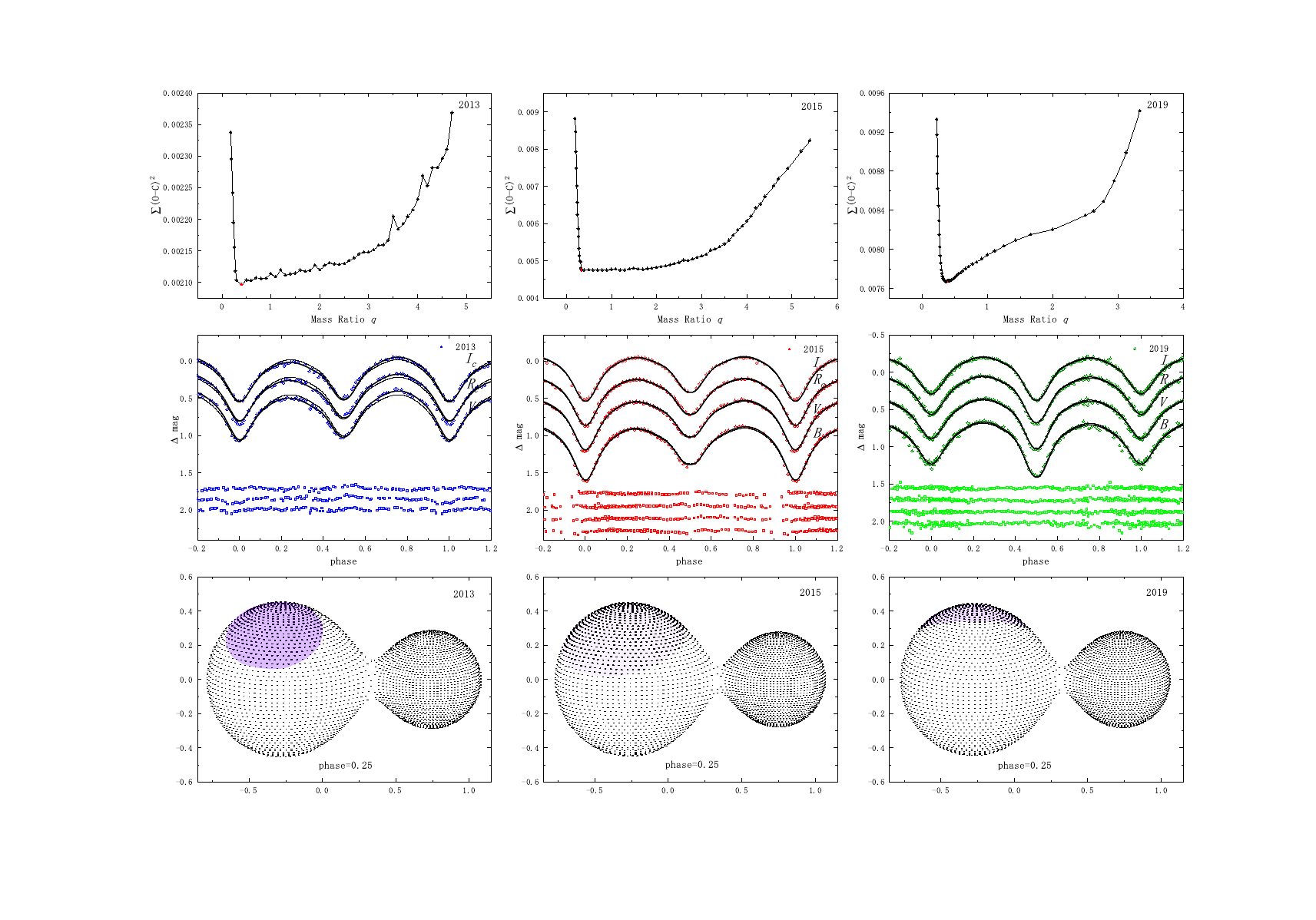}
 \end{center}
\caption{The upper panel: the q-search curves of V694 Peg for different years. The middle panel: the theoretical and observational light curves of V694 Peg. Dots are the observations. The solid lines represent the theoretical light curves with spots, and the dotted lines represent those without spots. The lower part of this panel provides the corresponding residues. The lower panel: the configuration of V694 Peg at 0.25 phase.}\label{fig:fit}
\end{figure*}

Since there was no reliable mass ratio q, q-search method was used to determine q. We had tried various models and found all convergence solutions are achieved at mode 3 (the contact model). The relation between the resulting sum $\Sigma$(\emph{O} - \emph{C})$^{2}$ of weighted square deviations and q in different years was shown in the upper panel of Figure \ref{fig:fit}. As can be seen from this figure, V694 Peg was A-type in 2013 and 2015, and then converted to W-type in 2019. The photometric solutions corresponding to the minimum weighted square deviations are taken as the initial photometric solutions. Then we made the mass ratio q as the adjustable parameter and obtained the final solutions through differential correction. The final solutions of the different years are listed in Table \ref{tab:6} and the corresponding theoretical light curves are shown in the middle panel of Figure \ref{fig:fit} with dotted lines. It was noticed that the theoretical curves can not fit the observations well around maxima. Considering that the amplitudes of the light variation become larger for shorter wavelengths, and this system is composed of late type stars. So we add dark spot in the binary model to solve this problem. After a lot of experiments, the solutions with spot for each set of light curves were obtained and plotted with solid lines in the middle panel of Figure \ref{fig:fit}. From this figure, one can see that the fitting are improved a lot. The dark spot parameters are latitude ($\phi$), longitude ($\theta$), angular radius (r$_{s}$), and the temperature factor (\emph{T}$_{f}$ = \emph{T}$_{s}$/\emph{T}$_{p}$; \emph{T}$_{f}$ is the ratio between the spot temperature \emph{T}$_{s}$ and the local effective temperature \emph{T}$_{p}$ of the adjacent photosphere), which are listed in Table \ref{tab:6}. The spot on the more-massive star in different years are shown in the lower panel of Figure \ref{fig:fit}. The theoretical color index curves are plotted in Figure \ref{CC}, and it can be seen that the theoretical curves is in good agreement with the observed curves. The results of our light curve analysis show that V694 Peg is an shallow contact binary and changes from A-type to W-type. It should be noted that the definition of W-type adopted here is that the less-massive star is hotter than the more-massive star, which is different from \cite{Binnendijk1970}'s definition by light curve. The three mass ratios obtained are very close. This means that the reliable photometric mass ratio for this system is around 0.36. Our results show that V694 Peg is of A-type in 2013, which is different from \citet{2016AJ....152...57D}'s. \citet{2016AJ....152...57D}'s model shows that V694 Peg belongs to the W-type, with the large-scale dark spots on the surface of the less-massive star. Other parameters are basically consistent with our results. This indicates that his model of W-type is greatly affected by spot, and the large-scale dark spots need to be added in a specific location to obtain a better fit. This also shows that spot have a great influence on the sub-type. Different spot parameters used in the modeling process would lead to different results.
\begin{deluxetable*}{cccccc}
\tabletypesize{\scriptsize}
\tablecaption{Photometric Solutions for V694 Peg.\label{tab:6}}
\tablewidth{0pt}
\tablehead{
\colhead{Parameters} & \colhead{2013 data solutions} & \colhead{2015 data solutions} & \colhead{2019 data solutions} & \colhead{2019 data solutions }& \colhead{2019 data solutions }\\
&&&&(q-search with spot&(q-search with spot
\\
&&&&on the primary component)&on the secondary component)
\\}
\decimalcolnumbers
\startdata
sub-type	&	A	&	A	&	W	&	A	&	A	\\
\emph{i}(deg) 	&	78.2(18)	&	80.22(37)	&	76.42(20)	&	78.30(38)	&	78.31(30)	\\
\emph{q}       	&	0.361(46)	&	0.3500(23)	&	0.370(34)	&	0.3717(66)	&	0.3700(23)	\\
      \emph{T}$_1$ (K) 	&	4450	&	4450	&	4450	&	4450	&	4450	\\
      \emph{T}$_2$ (K)/\emph{T}$_1$ (K)	&	0.9885(62)	&	0.9542(20)	&	1.0464(19)	&	0.9800(15)	&	0.9798(15)	\\
      \emph{L}$_{1B}$/(L$_{1B}$+L$_{2B}$) 	&	-	&	0.79841(76)	&	0.6177(28)	&	0.7455(12)	&	0.74558(74)	\\
      \emph{L}$_{2B}$/(L$_{1B}$+L$_{2B}$) 	&	-	&	0.20159(76)	&	0.3823(28)	&	0.2545(12)	&	0.25442(74)	\\
      \emph{L}$_{1V}$/(L$_{1V}$+L$_{2V}$) 	&	0.7276(75)	&	0.78406(72)	&	0.6383(24)	&	0.7385(12)	&	0.73856(68)	\\
      \emph{L}$_{2V}$/(L$_{1V}$+L$_{2V}$) 	&	0.2724(75)	&	0.21594(72)	&	0.3617(24)	&	0.2615(12)	&	0.26144(68)	\\
      \emph{L}$_{1R}$/(L$_{1R}$+L$_{2R}$)	&	0.7248(75)	&	0.77357(70)	&	0.6526(21)	&	0.7334(12)	&	0.73344(66)	\\
      \emph{L}$_{2R}$/(L$_{1R}$+L$_{2R}$)	&	0.2752(75)	&	0.22643(70)	&	0.3474(21)	&	0.2666(12)	&	0.26656(66)	\\
      \emph{L}$_{1I}$/(L$_{1I}$+L$_{2I}$) 	&	0.7226(75)	&	0.76444(66)	&	0.6643(19)	&	0.7294(12)	&	0.72941(63)	\\
      \emph{L}$_{2I}$/(L$_{1I}$+L$_{2I}$) 	&	0.2774(75)	&	0.23556(66)	&	0.3357(19)	&	0.2706(12)	&	0.27059(63)	\\
      $\Omega$$_1$=$\Omega$$_2$ 	&	2.534(89)	&	2.5434(55)	&	2.5391(48)	&	2.591(13)	&	2.5786(30)	\\  
      \emph{f} 	&	0.28(40)	&	0.143(26)	&	0.116(49)	&	0.125(58)	&	0.166(13)	\\
      \emph{r}$_1$(pole) 	&	0.4525(96)	&	0.44992(86)	&	0.4445(27)	&	0.4443(14)	&	0.44620(42)	\\
      \emph{r}$_2$(pole) 	&	0.289(46)	&	0.2783(23)	&	0.28182(74)	&	0.2823(62)	&	0.2844(17)	\\
      \emph{r}$_1$(side) 	&	0.487(12)	&	0.4833(11)	&	0.4766(37)	&	0.4764(18)	&	0.47887(57)	\\
      \emph{r}$_2$(side) 	&	0.304(57)	&	0.2908(28)	&	0.29442(85)	&	0.2949(75)	&	0.2975(21)	\\
      \emph{r}$_1$(back) 	&	0.518(11)	&	0.5115(14)	&	0.5049(51)	&	0.5048(18)	&	0.50789(79)	\\
      \emph{r}$_2$(back) 	&	0.35(11)	&	0.3281(51)	&	0.3309(13)	&	0.332(13)	&	0.3361(39)	\\
      $\phi$ (deg) 	&	0.498 	&	0.335 	&	0.432 	&	0.351 	&	0.428 	\\
      $\theta$ (deg)	&	4.447 	&	4.353 	&	1.595 	&	5.928 	&	2.474 	\\
      \emph{r}$_s$	&	0.687 	&	0.981 	&	0.906 	&	0.971 	&	0.890 	\\
      \emph{T}$_h$/\emph{T}$_p$ 	&	0.902 	&	0.984 	&	0.982 	&	0.974 	&	0.902 	\\
      $\rho$	&	0.0371 	&	0.0096 	&	0.0102 	&	0.0156 	&	0.0481 	\\
      $\sigma$(O-C)$^2$ 	&	9.19E-04	&	3.96E-03	&	6.44E-03	&	6.81E-03	&	6.37E-03	\\
\enddata
\tablecomments{$f$ is the fill-out factor. $\rho$ is the dark spot factor which are applies only to the same sphere.}
\end{deluxetable*}

\begin{figure*}
 \begin{center}
  \includegraphics[width=8cm]{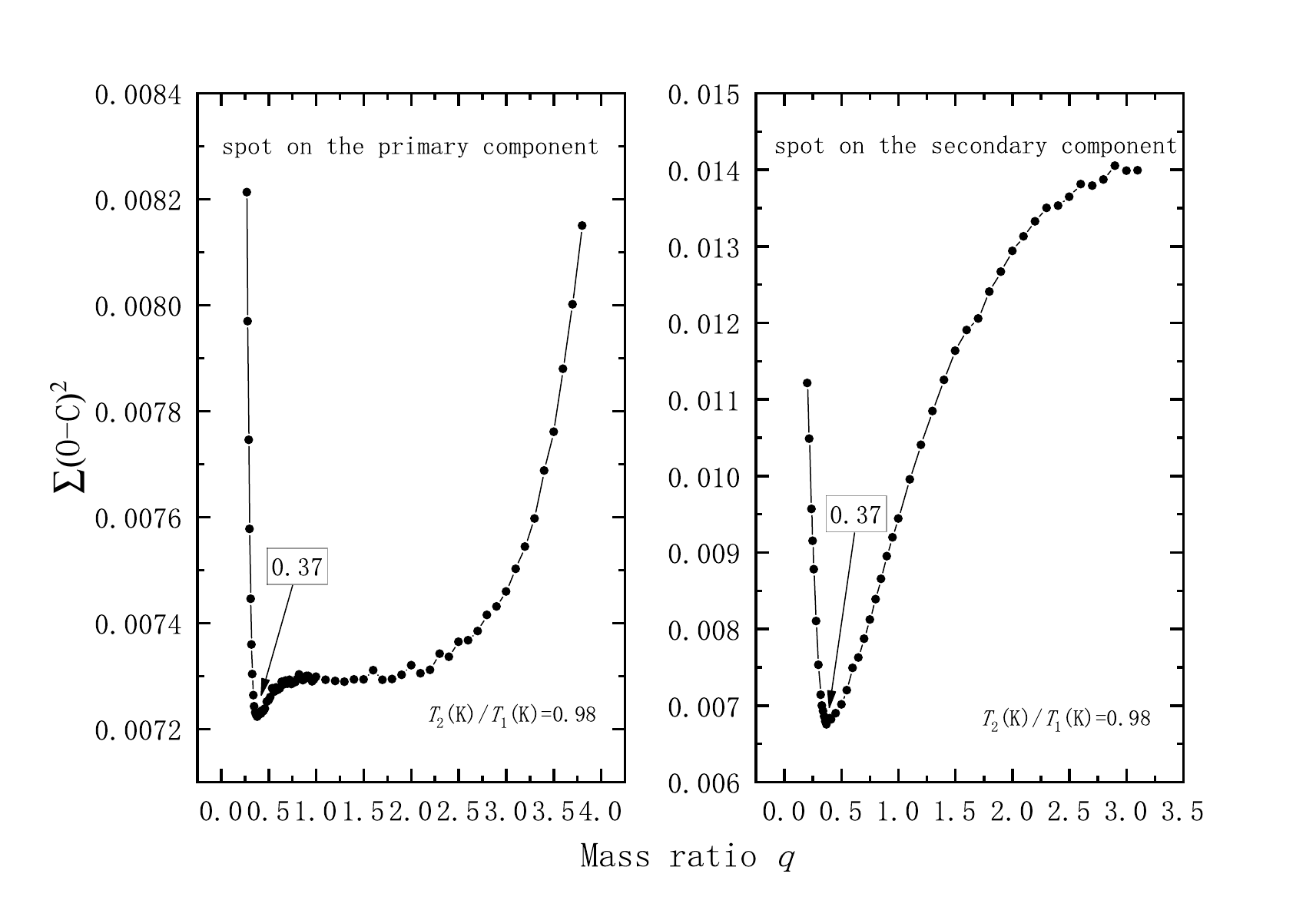}
 \end{center}
\caption{q-search graph with spot in 2019.}\label{qs}
\end{figure*}

From Figure \ref{fig:fit}, it can be noted that the W-D results show that the phase=0 in 2019 corresponds to the secondary minimum, which is different from 2013 and 2015. This is inconsistent with the epoch calculation. Since the phase of a non-totally eclipse binary system cannot be determined directly from the light curve, it is usually determined by the solutions of the W-D method. This is generally not confusing for a single set of light curves. However, for multiple sets of light curves, and involving the change between two sub-types, it needs to be explained. The phases are determined by the W-D results in this paper, and the phase=0 corresponds to the position of the more-massive star in front of the less-massive star. Therefore, we speculate that the sub-type of V694 Peg did not physically change in 2019, but rather that something caused the system to become a look-alike W-type system morphologically. In order to investigate the conversion between the two sub-type of V694 Peg, we try to study the influence of the dark spots, which are caused by the magnetic activity of late-type stars, on the two sub-types. As the photometric mass ratio of the three solutions is around 0.36, we fixed the mass ratio as 0.36 and added a dark spot to find the convergence solution. After a series of experiments, a convergent solution with a dark spot on the primary star was found. The corresponding spot parameters are listed in the columns 5 of Table \ref{tab:6}. Then fixed the parameters of the spot and added to the q-search process in 2019. The q-search process is the same as the previous approach. Finally, surprisingly, the mass ratio is found at exactly 0.37, which means the system could be A-type in 2019. Its W-type phenomena is caused by the spot due to the magnetic activity. We also try q-search with spots on the secondary star and got almost the same result. The q-search diagram is shown in Figure \ref{qs}, and the photometric solutions are listed in the last two columns of Table \ref{tab:6}. According to our spot model test, It is found that the influence of spot on sub-type of J2118 is dominant.
It is also important to note that W-D code often relate the temperature to average surface brightness, and temperature is affected by limb darkening, so it is difficult to completely rule out the effect of limb darkening on the results. This requires further work.

\section{DISCUSSIONS AND CONCLUSIONS}

\subsection{Absolute parameters of V694 Peg}
The three-sets multi-color light curves of V694 Peg have been analyzed by using WD code. Our results show that V694 Peg is a shallow contact binary converting from A-type to W-type. The asymmetry in the light curves could be explained by the spots model. Combined the photometric solutions and the parallax $Gpara$ given as 2.5337(459) mas by GAIA (\citealt{Gaia+2018}), we can calculate the absolute physical parameters of V694 Peg for further discussion. According to the formula of

\begin{equation}\label{10}
 (m - M)_v = 10 - 5lg \ Gpara ,
\end{equation}

\noindent the distance modulus of V694 Peg can be calculated. Adopted the latest apparent magnitude at maximum brightness as 14.79 mag of V694 Peg in \emph{V} band (\citealt{2016AJ....152...57D}), we can get the absolute magnitude in the \emph{V} band of V694 Peg. Together with the bolometric correction (e.g., \emph{BC}$_v$ = -0.58 mag for \emph{T} = 4450 K, \citealt{Worthey+2011}), the bolometric absolute magnitude of V694 Peg can be obtained by Eq. \ref{11}.
\begin{equation}\label{112}
  M_{bol} = M_v + BC_v.
\end{equation}

\noindent Thus, the total luminosity  \emph{L$_{T}$} of this system is obtained by Eq. \ref{12}.

\begin{equation}\label{12}
   L_{T} = L_{1} + L_{2} = 10^{-0.4(M_{bol} - 4.74)} * L_{\odot}
\end{equation}

\noindent According to Stefan-Boltzmann's law, we have
\begin{equation}\label{13}
  (L_1 + L_2)/L_{\odot} =(Ar_1)^2 * (\frac{T_1}{T_\odot})^4 + (Ar_2)^2 * (\frac{T_2}{T_\odot})^4,
\end{equation}
\noindent where \emph{T$_\odot$} represents the effective surface temperature of the sun, with a value of 5780 K. \emph{r$_1$} and \emph{r$_2$} represent the relative radii of the primary component and secondary component respectively, which can be calculated from the photometric solutions, i.e. \emph{r$_i$}=(\emph{r$_{ipole}$}*\emph{r$_{iside}$}*\emph{r$_{iback}$})$^{1/3}$. Since the photometric solutions given in different years are slightly different, all photometric parameters used in the calculation are the average values of the three sets of solutions.
Then, the semi-major axis \emph{A} is derived to be 1.5 $R_{\odot}$. Finally, according to Kepler's third law and mass ratio,
\begin{equation}\label{14}
  \frac{A^3}{P^2}= 74.5(M_1 + M_2),
\end{equation}
\begin{equation}\label{15}
  q=\frac{M_2}{M_1},
\end{equation}
\noindent two components masses \emph{M}$_1$ and \emph{M}$_2$ can be obtained. All the physical parameters are listed in Table \ref{pypa}, where the parameters with subscript 1 represent the parameters of the more-massive star. 

\begin{deluxetable*}{cccccccc}
\tabletypesize{\scriptsize}
\tablewidth{0pt} 
\tablecaption{Physical parameters of V694 Peg. \label{pypa}}
\tablehead{ \emph{M}$_{1}$(\emph{M}$_\odot$) & \emph{M}$_{2}$(\emph{M}$_\odot$) & \emph{R}$_{1}$(\emph{R}$_\odot$) & \emph{R}$_{2}$(\emph{R}$_\odot$) & \emph{L}$_{1}$ (\emph{L}$_\odot$)& \emph{L}$_{2}$(\emph{L}$_\odot$) & \emph{A}(\emph{R}$_\odot$) & \emph{D}(pc)} 
\colnumbers
\startdata 
 0.7(2)  & 0.3(1) & 0.7(1) & 0.5(1) & 0.2(1)  & 0.1(1) & 1.5(1) &  395(7) 
\enddata
\end{deluxetable*}
The absolute magnitude calculated by our method is 6.81 mag in 2013 and 6.75 mag in 2019. Another method of estimating absolute magnitude by period-color-luminosity was proposed by Rucinski \citep[e.g.][]{1994PASP..106..462R,1997PASP..109.1340R,2017AJ....154..125M}. The absolute magnitude of V694 Peg calculated by this method is 7.23 mag, which means that the system is over-luminosity in 2013 and 2019. 
\subsection{Light curve variation}
\begin{figure}
 \begin{center}
  \includegraphics[width=8cm]{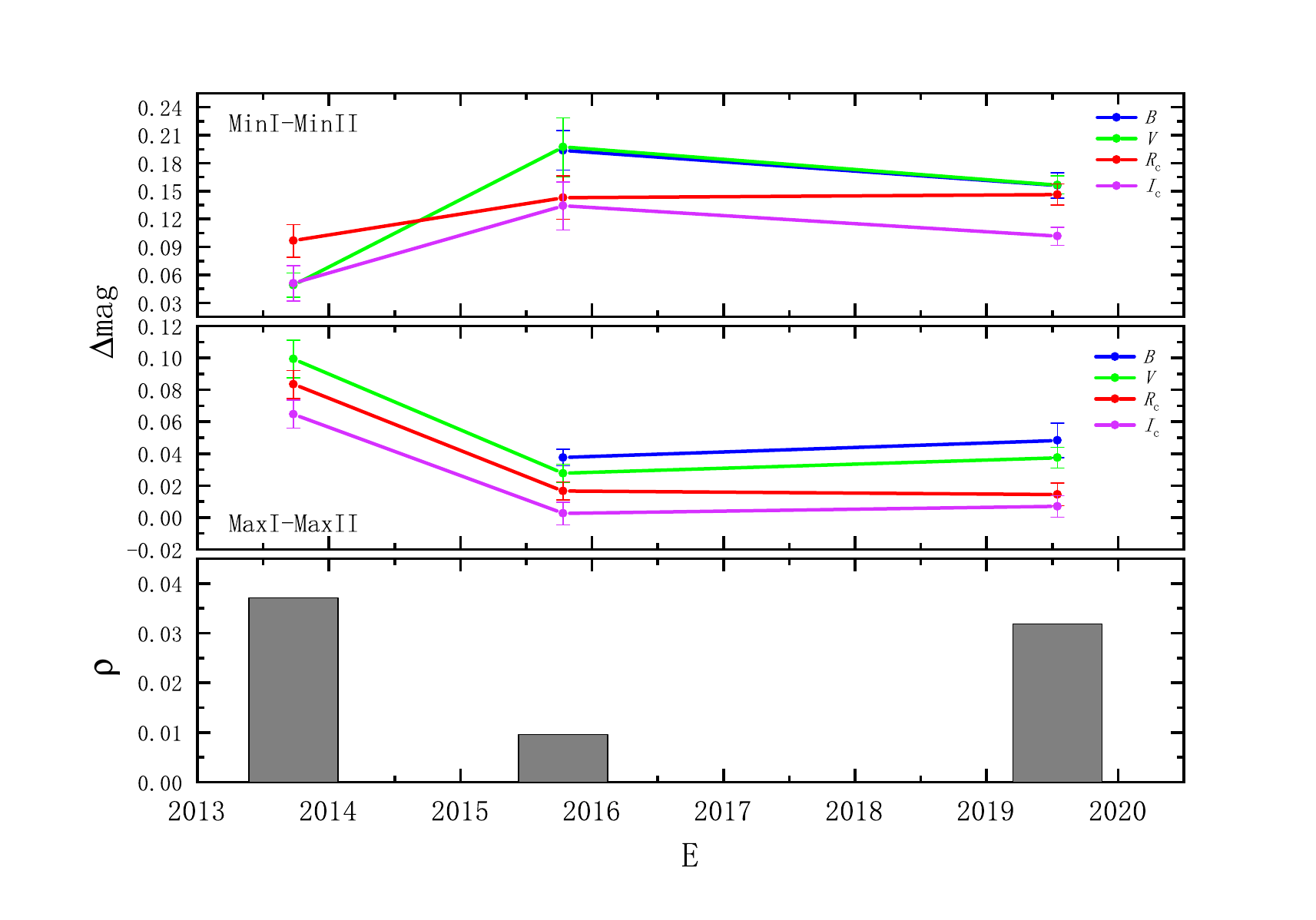}
 \end{center}
\caption{The upper panel shows the changes of the difference between two minima. The middle panel displays changes of the difference between the two maxima. The lower panel provides the dark spot factor variation.}\label{fig:13}
\end{figure}

Based on the light curves of V694 Peg in 2013, 2015, and 2019, We have found that the light curves of this system gradually deviates from EW-type. Besides this, the O'Connell effect in the light curves are more obvious in the shorter wavelength, which could be attributed to the dark spot on the surface of the primary component caused by magnetic activity. The dark spot of V694 Peg has evolved over a period of several years. The intensity of magnetic activity could be described by the dark spot factor, i.e. $\rho=\Delta E_{h}/E_{toll}$. $\Delta E_{h}$ represents the reduction in radiation intensity caused by the dark spot, and $E_{toll}$ is the radiation intensity of the star surface. According to the Stefan-Boltzmann law, we can have  $\rho=[r_{s}/2\pi][1-(T_{f})^{4}]$. The changes of the dark spot factor, the differences of two minima and maxima are plotted in Figure \ref{fig:13} and listed in Table \ref{tab:6}. The spot factor reached the largest value in 2013 and the lowest in 2015, and began to increase again in 2019. Its changes inversely correlated with that of the minima differences. When the dark spot factor was large, the light curves were more like the EW-type light variation, and when the dark spot factor became small, the light curves gradually deviates from the EW-type variation. This implies the change in the light curves of V694 Peg, which may be caused by magnetic activity. When the difference of eclipse depth between the primary minimum and the secondary minimum of the light curve is the smallest, the magnetic activity is the strongest. 

After adding the dark spot model, we conducted q-search on the light curve of V694 Peg in 2019, and obtained the same A-type as in 2013 and 2015, which indicates that the transformation of V694 Peg into W-type in 2019 is caused by dark spot activity. This is consistent with the high levels of magnetic activity observed in all A- and W-type conversion systems, as shown in Table \ref{AW} (Section \ref{trans}). This is the first model to confirm that spots can lead to A- and W-type conversion.

\begin{figure*}
 \begin{center}
  \includegraphics[width=8cm]{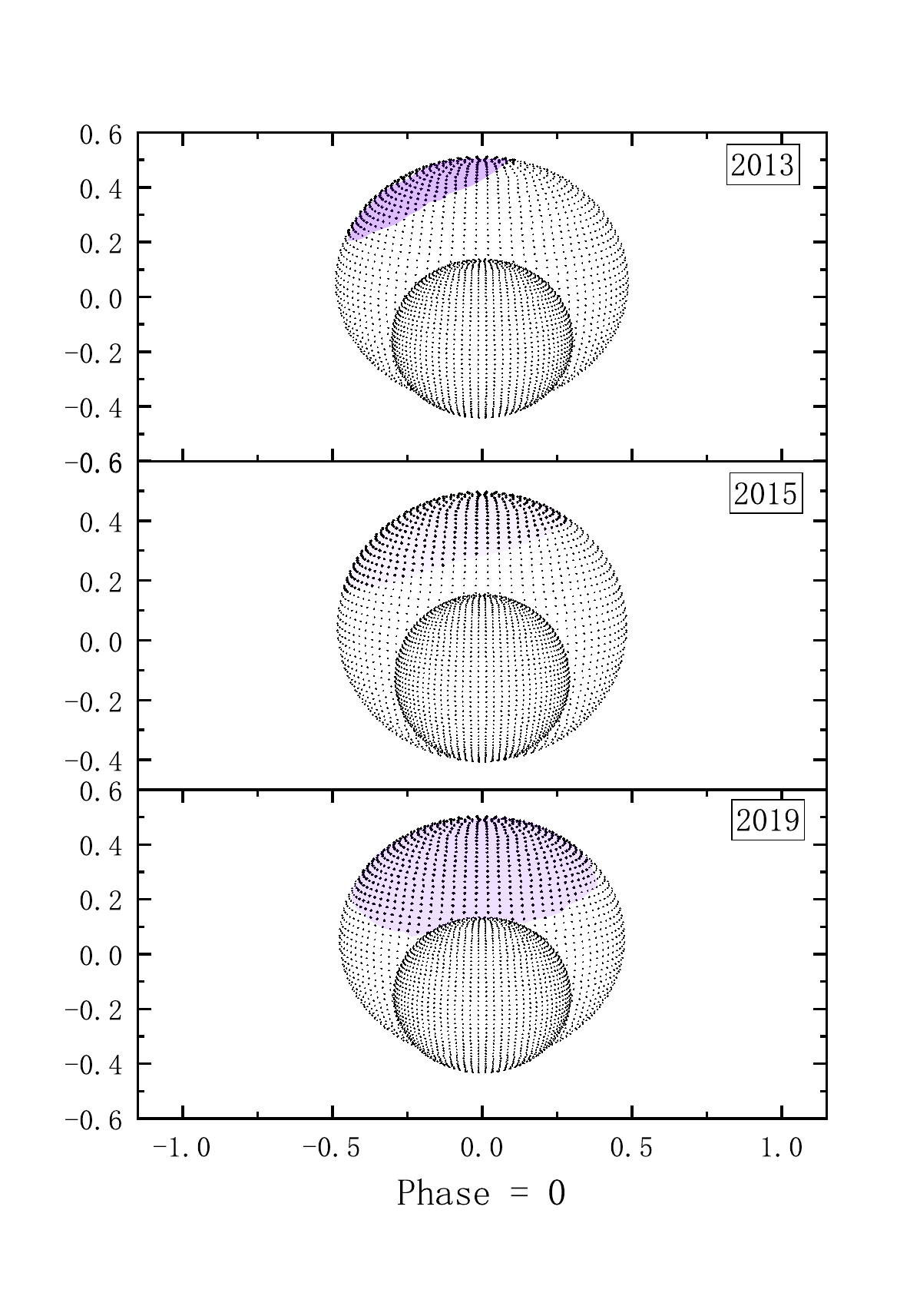}
 \end{center}
\caption{The graph of dark spot coverage at phase 0 of V694 Peg.}\label{GJDB}
\end{figure*}

\cite{Binnendijk1970} first classified binary stars into A- and W-type based on the feature of two light minima, but most binary stars could not be directly distinguished from their light curves. Later, other researchers physically divided the two sub-types through the structure characteristics of binary system obtained by W-D or other similar methods. In most cases, the results of the sub-types distinguished by the light curve are in agreement with those given by the W-D method. However, for some systems with special magnetic activity, inconsistency may occur. e.g., V694 Peg, spot activity affects not only the maximum but also the minimum of the light curves, which makes the light curve deviates from the EW-type. The dark spot lead to the change of the depth difference between the primary and secondary eclipses. We showed the configuration diagrams of V694 Peg at phase 0 in 2013, 2015 and 2019 (q-search with on the primary component) in Figure \ref{GJDB}. From Figure \ref{GJDB}, when the sub-type transition occurs in 2019, the dark spot is closer to the phase 0. However, the dark spot temperature in 2019 was not the lowest, which means that the location of the spots may be more important for the sub-type than the temperature of the spots. This may also be why magnetic activity is common in late-type contact binaries, but A-type and W-type transitions are not.

\begin{figure*}
 \begin{center}
  \includegraphics[width=10cm]{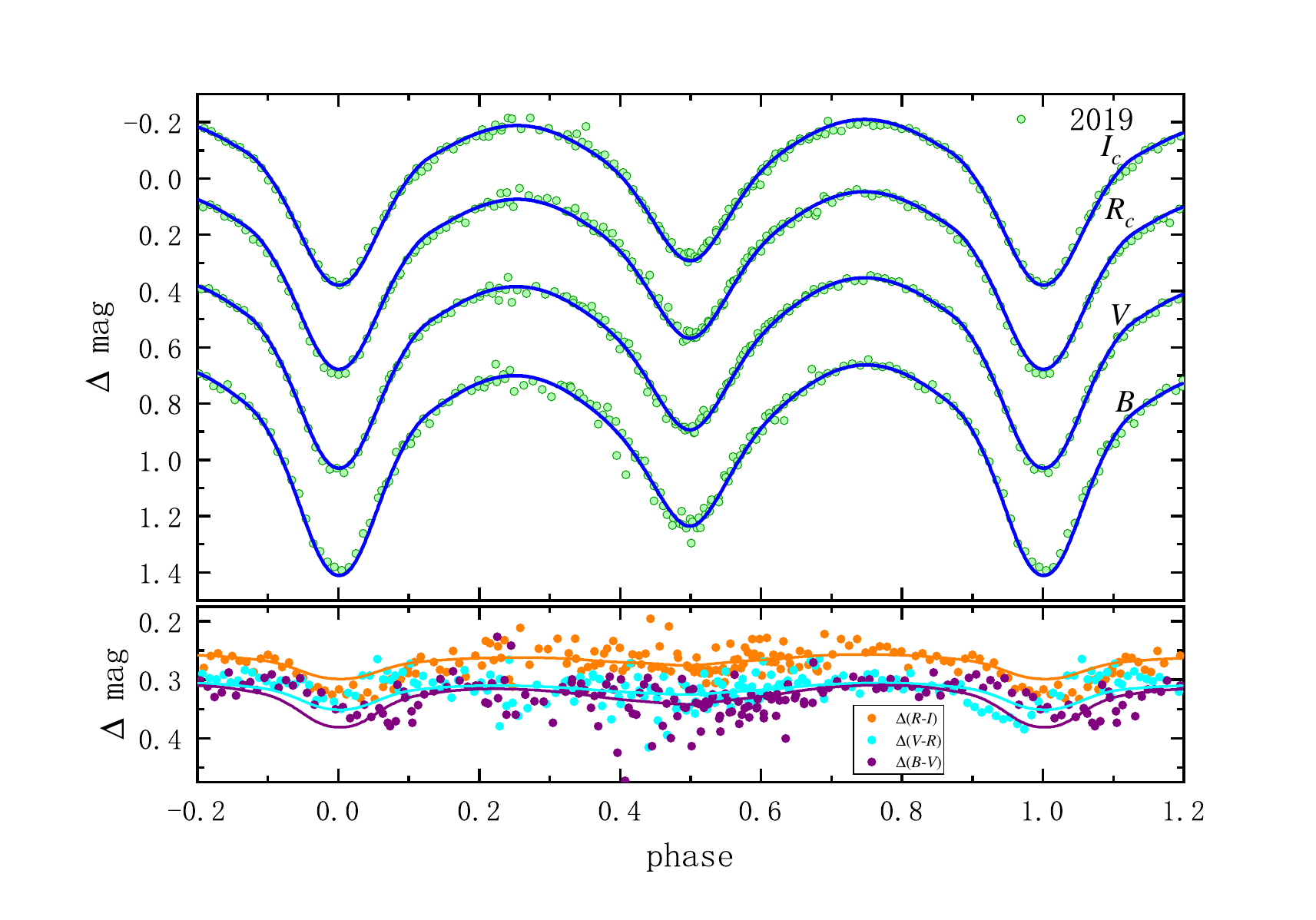}
 \end{center}
\caption{The upper panel: the theoretical and observational light curves of V694 Peg. Dots are the observations, and the solid line represents the theoretical light curve fitted at a single temperature. The lower panel: The color index curves of the V694 Peg in 2019. The solid points represent the observed color index, and the solid lines represent the theoretical color index curves.}\label{LCspot}
\end{figure*}

The effect of spots on sub-types can also be explained from another aspect. In essence, the W effect is a deviation from the strict Lucy model. A single temperature can be set for the system, as in the strict Roche model, and use spots on the primary component to account for any deviations. This was done for the complex MEM (the maximum entropy method) spot fitting for VW Cep by \cite{2000ApJ...531..467H}. Considering the difficulty of W-D code in simulating complex spots, especially when there are more than one, an approximate discussion is made. First, the mass ratio and related parameters are selected when the result is the solutions of W-type. the system temperature is set to a single temperature, and then two spots are added to the primary component to refit the light curves of 2019. One of the spot completely covers the primary component's surface to explain the deviation in apparent temperature; The other is to explain a small asymmetry in the light curves. The fitting results are shown in Figure \ref{LCspot}, and the temperature factor of the two spots is 0.95 and 0.98, respectively. The final results show that this method can also well fit the light curves, so as to explain that the deviation of the W-type phenomenon from the strict Lucy model can be caused by spots. 
\subsection{Period variation}

We have calculated all the times of light minimum of V694 Peg based on the data observed in 2013, 2015 and 2019, and the data obtained by the SuperWASP. A total of 505 minimum times spanning 17 years are used to construct the O-C curve and study the orbital period variation of this system. Through the O-C analysis, there is a long-term increase at a rate of $dP/dt$ = 4.3($\pm$0.3)$\times$ 10$^{-9}$ d yr$^{-1}$ and a periodic variation in the orbital period of V694 Peg. For the long-term period increase, the reasonable explanation is that the material of the less-massive component is transferring to the more-massive component, which leads to the orbital period increase, thus the system will evolve towards the separation direction. In the case of conservation of angular momentum, the mass transfer rate can be calculated according to the following equation,
\begin{equation}\label{6}
   \frac{\dot{P}}{P}=3(\frac{M_2}{M_1}-1)\frac{\dot{M_2}}{M_2}
\end{equation}
\noindent where \emph M$_1$ and \emph M$_2$ have been derived from the previous section. therefore, it was obtained as 2.4($\pm$ 1.6)$\times$10$^{-9}$\emph M$_\odot$yr$^{-1}$.

For the periodic variation, as the late-type nature of V694 Peg, one possible explanation is the results of the magnetic activity cycles of either of the components (Applegate mechanism, \citealt{1992ApJ..385..621A}).  According to the following equation given by  \citet{2000A&A...354..904R},
\begin{equation}\label{7}
   \Delta P = \sqrt{2(1-cos \omega)} * K
\end{equation}
\noindent The rate of periodic change $\Delta$ \emph P/\emph P can be derived. where \emph K is the semi-amplitude, $\omega$ is the oscillation period, given by Eq. \ref{P}. To produce such a change rate, the change in the required quadrupole momentum can be calculated using equation (\citealt{2002AN....323..424L}):
\begin{equation}\label{8}
   \Delta P/P= -9 \frac{\Delta Q_{1,2}}{M_{1,2}A^2}
\end{equation}
where \emph M$_{1,2}$ is the mass of each component. Thus, we obtain the
required quadrupole momentum variations of $\Delta$ Q$_1$ $\sim$ 3.4 $\times$ 10$^{46}$ g \ cm$^2$ and $\Delta$ Q$_2$ $\sim$ 1.2 $\times$ 10$^{46}$ g \ cm$^2$.

\begin{figure}
 \begin{center}
  \includegraphics[width=8cm]{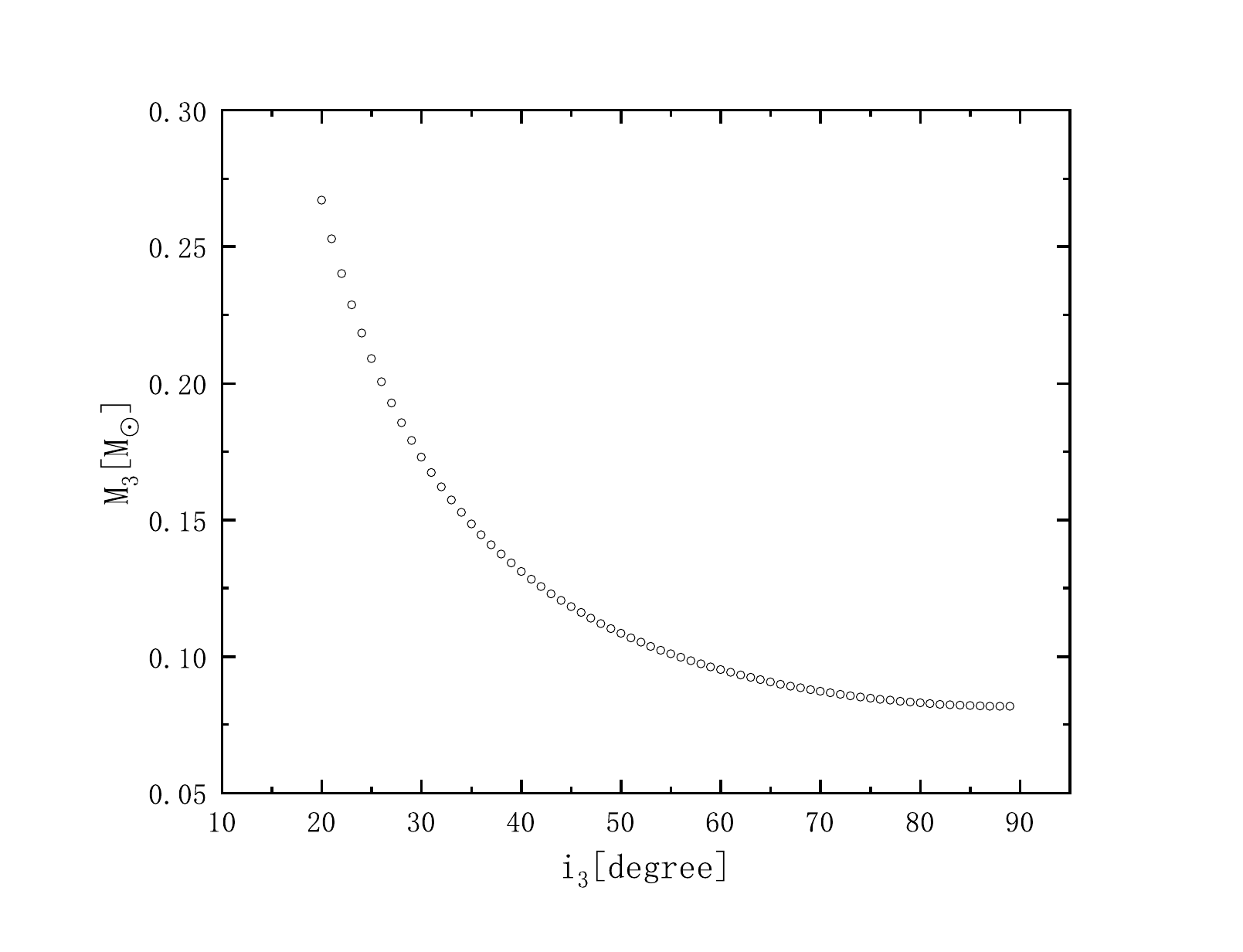}
 \end{center}
\caption{The relationship between the mass of third body and orbital inclination.}\label{fig:12}
\end{figure}

The other popular explanation of the periodic variation detected in the O-C curves is the light travel time effect caused by a third body (\citealt{Liao2010, 2019MNRAS.489.2677Z}). The third body orbiting around the contact binaries is believed to be one way that can cause their formation by removing angular momentum from the central pair at their early evolutionary stage (\citealt{Qian+2015b}). So the third body maybe the plausible reason of the cyclic period changes of V694 Peg. The mass function of the third body can be calculated by the following equation,
\begin{equation}\label{11}
   f(m)=\frac{4\pi^2}{GT^2}* (a'_{12}sin \ i_3)^3=\frac{(M_3sin \ i_3)^3}{(M_1+M_2+M_3)^2}
\end{equation}
Here, \emph M$_3$ is the mass of the third body. We plotted the relationship between the orbital inclination and the mass of the third body in Figure \ref{fig:12}. The parameters of the third body are listed in Table \ref{tab:tbody}. The minimum mass of the third body is 0.08\emph M$_\odot$ ($i_3$=90$^{\circ}$). We have tried to detect the third light during the light curves analysis, but failed. This may suggest that the contribution of the third light to the total luminosity is too small to detect.

\begin{deluxetable*}{lr}
\tablecaption{Parameters of the third component for V694 Peg.\label{tab:tbody}}
\tablewidth{0pt}
\tablehead{ Parameters &  Values  }
\decimalcolnumbers
\startdata
      $\omega$ [$^{\circ}$] & 0.01877(9)  \\
      $ K $ [d] & 0.002432(1) \\
      $P_{3}$ [yr]  & 11.81(6) \\
      $a_{12}$sin$i$ [AU] & 0.366(2)   \\
      $f$(m) ($M_\odot$) & 0.0005354(6)  \\
      $M_{3}(i_{3}=90^{\circ}$) ($M_\odot$) & 0.08(1) \\
      $a_{3}(i_{3}=90^{\circ}$) [AU] & 4.8(7) \\
\enddata
\end{deluxetable*}

\subsection{The transformation between two sub-types}\label{trans}

The nature of the two sub-types of contact binaries is still an open question as well as their evolutionary relationship. Our photometric solutions show that V694 Peg has converted from A-type to W-type if we haven't considered the spots effect. The detailed study of the systems transformed between the two sub-types have important role to study this issue. Up to now, only a few contact binaries have been reported converting between A-type and W-type. We have collected them and listed their parameters in Table \ref{AW}.
\begin{deluxetable*}{ccccccccccccccc}
\tabletypesize{\scriptsize}
\tablecaption{The parameters of the systems converted between A-type and W-type.\label{AW}}
\setlength{\tabcolsep}{0.3mm}{
\tablehead{
          \colhead{name}	&\colhead{$P$}	& \colhead{O'Connell effect} & type: $\rho$ & \colhead{$dP/dt(x10^{-9}$)}  & \colhead{$P_{3}$} &  \colhead{$K$} &	\colhead{q}	&	\colhead{$f$}	&	\colhead{$i$}	&	\colhead{$M_{1}$} 	&	\colhead{$M_{2}$}	& spectral type	&	type & References 	\\
      & \colhead{[d]} & & & \colhead{[d/yr]} & \colhead{[yr]} & \colhead{[d]} &  &  &\colhead{[$^{\circ}$]} &\colhead{[$M_\odot$]} &\colhead{[$M_\odot$]}&&&}
\decimalcolnumbers
\startdata
     V839 Oph &	0.409004 &Y &A:0;W:-0.00555	& 315 & 16.99 & 0.0075	&	 0.31	&	0.52 &  80 &	 1.6	&	0.5	&	F7V&	A-W	 &  16,13,14,7 \\
      AM Leo	&	0.365798&Y	&W:0;A:-0.02988 &-	&	50.5	&	0.0072	&	0.43	&	0.15	&	87	&	1.3	&	0.5	&F8V&	W-A	& 3,9,10,23 \\
      AH Cnc & 0.360458&Y & A:0;A:0.00283;W:0.005185;A:0.00273 & 4.29	&	35.26	&	0.061 &	0.15	&	0.51	&	83	&	1.2	&	0.2	&	 F7V &	W-A	&  2,6,15,22 \\
      AC Boo	&	0.352448&Y & A:-0.00277;W:-0.03161	& 267	&	 72.4 	&	0.047	&
	0.31	&	0.91	&	83	&	1.7	&	0.5	& F8V &	A-W &  1,4,17,18,21	\\
      RZ Com	&	0.338507&Y &W:0;A:$^*$;W:0	&39.7	&	45.1 & 	0.0065 &	0.44	&	0.07	&	86	&	1.1	&	0.5	&		K0&	W-A-W &	8,15,26 \\
      FG Hya 	&	0.327832&Y & A:0; W:0.00431	& -	&	54.44	&	0.0028	&	0.11	&	0.86	&	82	&	1.4	&	0.2	&G0&	A-W-A & 5,11,24	\\
      V1848 Ori	&	0.266350&Y & A:0.01380;A:0	& -	&	10.57	&	0.005	&	0.76	&	0.06	&	76	&	0.9	&	0.7	&	K2&	A-W	 &  19,20,25 \\
      V694 Peg &	0.224843& Y	& A:0.03713;A:0.00965;W:0.01017  & 4.3 & 11.81 & 0.0024	&	0.36	&	0.18 &  78 &	0.7	&	0.2	&K3-K5&	A-W	 &  this paper \\
\enddata
\tablecomments{ "$^*$": The light curves were not symmetrical, but the author did not add the spot to fit; minus "-" (The 4th column): hot spot.\\
References: 1.\citealt{1983AAS..52..463}; 2.\citealt{1984AAS..58..405}; 3.\citealt{1991IBVS.3630....1D}; 4.\citealt{1993ASPC..38..269};
    5.\citealt{yang2000};
    6.\citealt{2003AJ..126..2954};
    7.\citealt{2003ApSS.288..259P};
    8.\citealt{2004NA..9..273};
    9.\citealt{2004PASP..116..337H};
    10.\citealt{2005AJ....129.1686Q};
    11.\citealt{Qian+Yang2005};  12.\citealt{2006AJ....131.3028Q};
    13.\citealt{2006ApSS.304..125G}; 14.\citealt{2006ApSS.304...35S};
    15.\citealt{2008ChJAA...8..465H};
    16.\citealt{2010JKAS...43..201H};
    17.\citealt{2010IBVS..5951..1}; 18.\citealt{2010JAVSO..38..57};
    19.\citealt{2012IBVS.6012....1P}; 20.\citealt{2014NewA...28...23K}; 21.\citealt{2015IBVS.6142....1N}; 22.\citealt{2016RAA..16..157}; 23.\citealt{2017IBVS.6227....1G}; 24.\citealt{2017RAA....17..128Z}; 25.\citealt{2020OAst...29...72D};. 26.\citealt{2005PASJ...57..977Q}. }}
\end{deluxetable*}

From Table \ref{AW}, we can see that all the systems belong to late-type binaries with a spectral type range of K5 to F7 and a period range of 0.22 - 0.42 days. V694 Peg recorded as the latest one with the lowest orbital period, which has the fastest rotation speed, deep convection, and the most significant dark spot activity. All the systems show the O'Connell effect with unequal height maximum on the light curves, meaning that they have high level magnetic activity and are consistent with their late-type properties. Among these systems, Four of them have been detected having long-term period increase, and the others keep period constant. As the late-type binaries are active and drive magnetized wind, they lose angular momentum via magnetic wind and cause orbital period shrinking. All these converting systems show orbital period increase or constant, which implies that they all have mass transfer from the less-massive component to the more-massive one. The cyclic period variations has been detected in all these systems suggesting that there are possible third body or the magnetic cycles. The oscillation periods could be 7 years to 70 years. In addition, the mass ratios and contact degree of these systems are not concentrated within a specific range. That means the transformation between A-type and W-type is not correlated with mass ratio and the thickness of the common convective envelope of the contact binaries.

In conclusion, the contact binary converting between the two sub-types may be correlated with their late-type nature and the mass transfer from the less-massive star to the more-massive one, or even the existence of the third companion. For investigating the relation of A-type and W-type phenomena, discovery of more converting contact binaries are essential. With the release of data from surveys such as the Transiting Exoplanet Survey Satellite \citep[TESS,][]{2015JATIS...1a4003R}, which has 27.4 days of continuous observations, it is expected to find more systems that switch between A- and W-type. Thus, the cause of the W-type phenomenon is further revealed.

\section*{Acknowledgements}
Special thanks to Prof. Chris Koen for the light curves data in 2015 and Prof. Andrew Norton for the SuperWasp Sky Survey data. Thanks also to Prof. R.E.Wilson for his valuable advice on the use of WD code. This work was supported by the National Natural Science Foundation of China (No. 11922306), by the Yunnan Fundamental Research Projects (No. 202401CF070039) and Chinese Academy of Sciences (CAS) Interdisciplinary Innovation Team. We acknowledge the use of 85 cm telescope in Xinglong station of National Astronomical Observations, Chinese Academy of Sciences.

\section*{Data Availability}

The data underlying this article are available in the article and in its online supplementary material.

\appendix
\section{Some extra material}
The data of \emph{B}\emph{V}\emph{R}$_{c}$\emph{I}$_{c}$ light curves of V694 Peg\\
table 4 -continue All available times of minimum

\bibliography{sample631}{}

\begin{thebibliography}{}
\expandafter\ifx\csname natexlab\endcsname\relax\def\natexlab#1{#1}\fi
\providecommand{\url}[1]{\href{#1}{#1}}
\providecommand{\dodoi}[1]{doi:~\href{http://doi.org/#1}{\nolinkurl{#1}}}
\providecommand{\doeprint}[1]{\href{http://ascl.net/#1}{\nolinkurl{http://ascl.net/#1}}}
\providecommand{\doarXiv}[1]{\href{https://arxiv.org/abs/#1}{\nolinkurl{https://arxiv.org/abs/#1}}}

\bibitem[{{Alton}(2010)}]{2010JAVSO..38..57}
{Alton}, K.~B. 2010, \jaavso, 38, 57

\bibitem[{{Applegate}(1992)}]{1992ApJ..385..621A}
{Applegate}, J.~H. 1992, \apj, 385, 621, \dodoi{10.1086/170967}

\bibitem[{{Binnendijk}(1970)}]{Binnendijk1970}
{Binnendijk}, L. 1970, Vistas in Astronomy, 12, 217, \dodoi{10.1016/0083-6656(70)90041-3}

\bibitem[{{{\c{S}}enavc{\i}} {et~al.}(2006){{\c{S}}enavc{\i}}, {Albayrak}, {Selam}, \& {Ak}}]{2006ApSS.304...35S}
{{\c{S}}enavc{\i}}, H.~V., {Albayrak}, B., {Selam}, S.~O., \& {Ak}, T. 2006, \apss, 304, 35, \dodoi{10.1007/s10509-006-9132-x}

\bibitem[{{Davoudi} {et~al.}(2020){Davoudi}, {Poro}, {Alicavus}, {Halavati}, {Doostmohammadi}, {Shahdadi}, {Vahedi}, {Pishahang}, {Zare}, {Jamali}, {Salajeghe}, {Jahediparizi}, {Ashta}, \& {Shojaatalhosseini}}]{2020OAst...29...72D}
{Davoudi}, F., {Poro}, A., {Alicavus}, F., {et~al.} 2020, Open Astronomy, 29, 72, \dodoi{10.1515/astro-2020-0013}

\bibitem[{{Derman} {et~al.}(1991){Derman}, {Demircan}, \& {Dundar}}]{1991IBVS.3630....1D}
{Derman}, E., {Demircan}, O., \& {Dundar}, H. 1991, Information Bulletin on Variable Stars, 3630, 1

\bibitem[{{Djura{\v{s}}evi{\'c}} {et~al.}(2016){Djura{\v{s}}evi{\'c}}, {Essam}, {Latkovi{\'c}}, {Cs{\'e}ki}, {El-Sadek}, {Abo-Elala}, \& {Hayman}}]{2016AJ....152...57D}
{Djura{\v{s}}evi{\'c}}, G., {Essam}, A., {Latkovi{\'c}}, O., {et~al.} 2016, \aj, 152, 57, \dodoi{10.3847/0004-6256/152/3/57}

\bibitem[{{Drake} {et~al.}(2014){Drake}, {Graham}, {Djorgovski}, {Catelan}, {Mahabal}, {Torrealba}, {Garc{\'\i}a-{\'A}lvarez}, {Donalek}, {Prieto}, {Williams}, {Larson}, {Christen sen}, {Belokurov}, {Koposov}, {Beshore}, {Boattini}, {Gibbs}, {Hill}, {Kowalski}, {Johnson}, \& {Shelly}}]{2014ApJS..213....9D}
{Drake}, A.~J., {Graham}, M.~J., {Djorgovski}, S.~G., {et~al.} 2014, \apjs, 213, 9, \dodoi{10.1088/0067-0049/213/1/9}

\bibitem[{{Eaton} {et~al.}(1980){Eaton}, {Wu}, \& {Rucinski}}]{Eaton1980}
{Eaton}, J.~A., {Wu}, C.~C., \& {Rucinski}, S.~M. 1980, \apj, 239, 919, \dodoi{10.1086/158179}

\bibitem[{{Eggen}(1967)}]{Eggen1967}
{Eggen}, O.~J. 1967, \memras, 70, 111

\bibitem[{{Gaia Collaboration} {et~al.}(2018){Gaia Collaboration}, {Brown}, {Vallenari}, {Prusti}, {de Bruijne}, {Babusiaux}, {Bailer-Jones}, {Biermann}, {Evans}, {Eyer}, {Jansen}, {Jordi}, {Klioner}, {Lammers}, {Lindegren}, {Luri}, {Mignard}, {Panem}, {Pourbaix}, {Randich}, {Sartoretti}, {Siddiqui}, {Soubiran}, {van Leeuwen}, {Walton}, {Arenou}, {Bastian}, {Cropper}, {Drimmel}, {Katz}, {Lattanzi}, {Bakker}, {Cacciari}, {Casta{\~n}eda}, {Chaoul}, {Cheek}, {De Angeli}, {Fabricius}, {Guerra}, {Holl}, {Masana}, {Messineo}, {Mowlavi}, {Nienartowicz}, {Panuzzo}, {Portell}, {Riello}, {Seabroke}, {Tanga}, {Th{\'e}venin}, {Gracia-Abril}, {Comoretto}, {Garcia-Reinaldos}, {Teyssier}, {Altmann}, {Andrae}, {Audard}, {Bellas-Velidis}, {Benson}, {Berthier}, {Blomme}, {Burgess}, {Busso}, {Carry}, {Cellino}, {Clementini}, {Clotet}, {Creevey}, {Davidson}, {De Ridder}, {Delchambre}, {Dell'Oro}, {Ducourant}, {Fern{\'a}ndez-Hern{\'a}ndez}, {Fouesneau}, {Fr{\'e}mat}, {Galluccio}, {Garc{\'\i}a-Torres},
  {Gonz{\'a}lez-N{\'u}{\~n}ez}, {Gonz{\'a}lez-Vidal}, {Gosset}, {Guy}, {Halbwachs}, {Hambly}, {Harrison}, {Hern{\'a}ndez}, {Hestroffer}, {Hodgkin}, {Hutton}, {Jasniewicz}, {Jean-Antoine-Piccolo}, {Jordan}, {Korn}, {Krone-Martins}, {Lanzafame}, {Lebzelter}, {L{\"o}ffler}, {Manteiga}, {Marrese}, {Mart{\'\i}n-Fleitas}, {Moitinho}, {Mora}, {Muinonen}, {Osinde}, {Pancino}, {Pauwels}, {Petit}, {Recio-Blanco}, {Richards}, {Rimoldini}, {Robin}, {Sarro}, {Siopis}, {Smith}, {Sozzetti}, {S{\"u}veges}, {Torra}, {van Reeven}, {Abbas}, {Abreu Aramburu}, {Accart}, {Aerts}, {Altavilla}, {{\'A}lvarez}, {Alvarez}, {Alves}, {Anderson}, {Andrei}, {Anglada Varela}, {Antiche}, {Antoja}, {Arcay}, {Astraatmadja}, {Bach}, {Baker}, {Balaguer-N{\'u}{\~n}ez}, {Balm}, {Barache}, {Barata}, {Barbato}, {Barblan}, {Barklem}, {Barrado}, {Barros}, {Barstow}, {Bartholom{\'e} Mu{\~n}oz}, {Bassilana}, {Becciani}, {Bellazzini}, {Berihuete}, {Bertone}, {Bianchi}, {Bienaym{\'e}}, {Blanco-Cuaresma}, {Boch}, {Boeche}, {Bombrun}, {Borrachero},
  {Bossini}, {Bouquillon}, {Bourda}, {Bragaglia}, {Bramante}, {Breddels}, {Bressan}, {Brouillet}, {Br{\"u}semeister}, {Brugaletta}, {Bucciarelli}, {Burlacu}, {Busonero}, {Butkevich}, {Buzzi}, {Caffau}, {Cancelliere}, {Cannizzaro}, {Cantat-Gaudin}, {Carballo}, {Carlucci}, {Carrasco}, {Casamiquela}, {Castellani}, {Castro-Ginard}, {Charlot}, {Chemin}, {Chiavassa}, {Cocozza}, {Costigan}, {Cowell}, {Crifo}, {Crosta}, {Crowley}, {Cuypers}, {Dafonte}, {Damerdji}, {Dapergolas}, {David}, {David}, {de Laverny}, {De Luise}, {De March}, {de Martino}, {de Souza}, {de Torres}, {Debosscher}, {del Pozo}, {Delbo}, {Delgado}, {Delgado}, {Di Matteo}, {Diakite}, {Diener}, {Distefano}, {Dolding}, {Drazinos}, {Dur{\'a}n}, {Edvardsson}, {Enke}, {Eriksson}, {Esquej}, {Eynard Bontemps}, {Fabre}, {Fabrizio}, {Faigler}, {Falc{\~a}o}, {Farr{\`a}s Casas}, {Federici}, {Fedorets}, {Fernique}, {Figueras}, {Filippi}, {Findeisen}, {Fonti}, {Fraile}, {Fraser}, {Fr{\'e}zouls}, {Gai}, {Galleti}, {Garabato}, {Garc{\'\i}a-Sedano}, {Garofalo},
  {Garralda}, {Gavel}, {Gavras}, {Gerssen}, {Geyer}, {Giacobbe}, {Gilmore}, {Girona}, {Giuffrida}, {Glass}, {Gomes}, {Granvik}, {Gueguen}, {Guerrier}, {Guiraud}, {Guti{\'e}rrez-S{\'a}nchez}, {Haigron}, {Hatzidimitriou}, {Hauser}, {Haywood}, {Heiter}, {Helmi}, {Heu}, {Hilger}, {Hobbs}, {Hofmann}, {Holland}, {Huckle}, {Hypki}, {Icardi}, {Jan{\ss}en}, {Jevardat de Fombelle}, {Jonker}, {Juh{\'a}sz}, {Julbe}, {Karampelas}, {Kewley}, {Klar}, {Kochoska}, {Kohley}, {Kolenberg}, {Kontizas}, {Kontizas}, {Koposov}, {Kordopatis}, {Kostrzewa-Rutkowska}, {Koubsky}, {Lambert}, {Lanza}, {Lasne}, {Lavigne}, {Le Fustec}, {Le Poncin-Lafitte}, {Lebreton}, {Leccia}, {Leclerc}, {Lecoeur-Taibi}, {Lenhardt}, {Leroux}, {Liao}, {Licata}, {Lindstr{\o}m}, {Lister}, {Livanou}, {Lobel}, {L{\'o}pez}, {Managau}, {Mann}, {Mantelet}, {Marchal}, {Marchant}, {Marconi}, {Marinoni}, {Marschalk{\'o}}, {Marshall}, {Martino}, {Marton}, {Mary}, {Massari}, {Matijevi{\v{c}}}, {Mazeh}, {McMillan}, {Messina}, {Michalik}, {Millar}, {Molina}, {Molinaro},
  {Moln{\'a}r}, {Montegriffo}, {Mor}, {Morbidelli}, {Morel}, {Morris}, {Mulone}, {Muraveva}, {Musella}, {Nelemans}, {Nicastro}, {Noval}, {O'Mullane}, {Ord{\'e}novic}, {Ord{\'o}{\~n}ez-Blanco}, {Osborne}, {Pagani}, {Pagano}, {Pailler}, {Palacin}, {Palaversa}, {Panahi}, {Pawlak}, {Piersimoni}, {Pineau}, {Plachy}, {Plum}, {Poggio}, {Poujoulet}, {Pr{\v{s}}a}, {Pulone}, {Racero}, {Ragaini}, {Rambaux}, {Ramos-Lerate}, {Regibo}, {Reyl{\'e}}, {Riclet}, {Ripepi}, {Riva}, {Rivard}, {Rixon}, {Roegiers}, {Roelens}, {Romero-G{\'o}mez}, {Rowell}, {Royer}, {Ruiz-Dern}, {Sadowski}, {Sagrist{\`a} Sell{\'e}s}, {Sahlmann}, {Salgado}, {Salguero}, {Sanna}, {Santana-Ros}, {Sarasso}, {Savietto}, {Schultheis}, {Sciacca}, {Segol}, {Segovia}, {S{\'e}gransan}, {Shih}, {Siltala}, {Silva}, {Smart}, {Smith}, {Solano}, {Solitro}, {Sordo}, {Soria Nieto}, {Souchay}, {Spagna}, {Spoto}, {Stampa}, {Steele}, {Steidelm{\"u}ller}, {Stephenson}, {Stoev}, {Suess}, {Surdej}, {Szabados}, {Szegedi-Elek}, {Tapiador}, {Taris}, {Tauran}, {Taylor},
  {Teixeira}, {Terrett}, {Teyssandier}, {Thuillot}, {Titarenko}, {Torra Clotet}, {Turon}, {Ulla}, {Utrilla}, {Uzzi}, {Vaillant}, {Valentini}, {Valette}, {van Elteren}, {Van Hemelryck}, {van Leeuwen}, {Vaschetto}, {Vecchiato}, {Veljanoski}, {Viala}, {Vicente}, {Vogt}, {von Essen}, {Voss}, {Votruba}, {Voutsinas}, {Walmsley}, {Weiler}, {Wertz}, {Wevers}, {Wyrzykowski}, {Yoldas}, {{\v{Z}}erjal}, {Ziaeepour}, {Zorec}, {Zschocke}, {Zucker}, {Zurbach}, \& {Zwitter}}]{Gaia+2018}
{Gaia Collaboration}, {Brown}, A.~G.~A., {Vallenari}, A., {et~al.} 2018, \aap, 616, A1, \dodoi{10.1051/0004-6361/201833051}

\bibitem[{{Gazeas} \& {St{\c{e}}pie{\'n}}(2008)}]{Gazeas2008}
{Gazeas}, K., \& {St{\c{e}}pie{\'n}}, K. 2008, \mnras, 390, 1577, \dodoi{10.1111/j.1365-2966.2008.13844.x}

\bibitem[{{Gazeas} \& {Niarchos}(2006)}]{Gazeas2006}
{Gazeas}, K.~D., \& {Niarchos}, P.~G. 2006, \mnras, 370, L29, \dodoi{10.1111/j.1745-3933.2006.00182.x}

\bibitem[{{Gazeas} {et~al.}(2006){Gazeas}, {Niarchos}, \& {Gradoula}}]{2006ApSS.304..125G}
{Gazeas}, K.~D., {Niarchos}, P.~G., \& {Gradoula}, G.~P. 2006, \apss, 304, 125, \dodoi{10.1007/s10509-006-9090-3}

\bibitem[{{Gorda} \& {Matveeva}(2017)}]{2017IBVS.6227....1G}
{Gorda}, S.~Y., \& {Matveeva}, E.~A. 2017, Information Bulletin on Variable Stars, 6227, 1, \dodoi{10.22444/IBVS.6227}

\bibitem[{{Graham} {et~al.}(2019){Graham}, {Kulkarni}, {Bellm}, {Adams}, {Barbarino}, {Blagorodnova}, {Bodewits}, {Bolin}, {Brady}, {Cenko}, {Chang}, {Coughlin}, {De}, {Eadie}, {Farnham}, {Feindt}, {Franckowiak}, {Fremling}, {Gezari}, {Ghosh}, {Goldstein}, {Golkhou}, {Goobar}, {Ho}, {Huppenkothen}, {Ivezi{\'c}}, {Jones}, {Juric}, {Kaplan}, {Kasliwal}, {Kelley}, {Kupfer}, {Lee}, {Lin}, {Lunnan}, {Mahabal}, {Miller}, {Ngeow}, {Nugent}, {Ofek}, {Prince}, {Rauch}, {van Roestel}, {Schulze}, {Singer}, {Sollerman}, {Taddia}, {Yan}, {Ye}, {Yu}, {Barlow}, {Bauer}, {Beck}, {Belicki}, {Biswas}, {Brinnel}, {Brooke}, {Bue}, {Bulla}, {Burruss}, {Connolly}, {Cromer}, {Cunningham}, {Dekany}, {Delacroix}, {Desai}, {Duev}, {Feeney}, {Flynn}, {Frederick}, {Gal-Yam}, {Giomi}, {Groom}, {Hacopians}, {Hale}, {Helou}, {Henning}, {Hover}, {Hillenbrand}, {Howell}, {Hung}, {Imel}, {Ip}, {Jackson}, {Kaspi}, {Kaye}, {Kowalski}, {Kramer}, {Kuhn}, {Landry}, {Laher}, {Mao}, {Masci}, {Monkewitz}, {Murphy}, {Nordin}, {Patterson}, {Penprase},
  {Porter}, {Rebbapragada}, {Reiley}, {Riddle}, {Rigault}, {Rodriguez}, {Rusholme}, {van Santen}, {Shupe}, {Smith}, {Soumagnac}, {Stein}, {Surace}, {Szkody}, {Terek}, {Van Sistine}, {van Velzen}, {Vestrand}, {Walters}, {Ward}, {Zhang}, \& {Zolkower}}]{2019PASP..131g8001G}
{Graham}, M.~J., {Kulkarni}, S.~R., {Bellm}, E.~C., {et~al.} 2019, \pasp, 131, 078001, \dodoi{10.1088/1538-3873/ab006c}

\bibitem[{{Hanna}(2010)}]{2010JKAS...43..201H}
{Hanna}, M.~A. 2010, Journal of Korean Astronomical Society, 43, 201, \dodoi{10.5303/JKAS.2010.43.6.201}

\bibitem[{{He} \& {Qian}(2008)}]{2008ChJAA...8..465H}
{He}, J.-J., \& {Qian}, S.-B. 2008, \cjaa, 8, 465, \dodoi{10.1088/1009-9271/8/4/10}

\bibitem[{{Hendry} \& {Mochnacki}(2000)}]{2000ApJ...531..467H}
{Hendry}, P.~D., \& {Mochnacki}, S.~W. 2000, \apj, 531, 467, \dodoi{10.1086/308427}

\bibitem[{{Hiller} {et~al.}(2004){Hiller}, {Osborn}, \& {Terrell}}]{2004PASP..116..337H}
{Hiller}, M.~E., {Osborn}, W., \& {Terrell}, D. 2004, \pasp, 116, 337, \dodoi{10.1086/383098}

\bibitem[{{Hrivnak}(1993)}]{1993ASPC..38..269}
{Hrivnak}, B.~J. 1993, in Astronomical Society of the Pacific Conference Series, Vol.~38, New Frontiers in Binary Star Research, ed. K.-C. {Leung} \& I.-S. {Nha}, 269

\bibitem[{{Jabbar} \& {Kopal}(1983)}]{1983Ap&SS..92...99J}
{Jabbar}, S.~R., \& {Kopal}, Z. 1983, \apss, 92, 99, \dodoi{10.1007/BF00653589}

\bibitem[{{Kochanek} {et~al.}(2017){Kochanek}, {Shappee}, {Stanek}, {Holoien}, {Thompson}, {Prieto}, {Dong}, {Shields}, {Will}, {Britt}, {Perzanowski}, \& {Pojma{\'n}ski}}]{2017PASP..129j4502K}
{Kochanek}, C.~S., {Shappee}, B.~J., {Stanek}, K.~Z., {et~al.} 2017, \pasp, 129, 104502, \dodoi{10.1088/1538-3873/aa80d9}

\bibitem[{{Koen} {et~al.}(2016){Koen}, {Koen}, \& {Gray}}]{Koen2016}
{Koen}, C., {Koen}, T., \& {Gray}, R.~O. 2016, \aj, 151, 168, \dodoi{10.3847/0004-6256/151/6/168}

\bibitem[{{Kriwattanawong} \& {Poojon}(2014)}]{2014NewA...28...23K}
{Kriwattanawong}, W., \& {Poojon}, P. 2014, \na, 28, 23, \dodoi{10.1016/j.newast.2013.09.009}

\bibitem[{{Lanza} \& {Rodon{\`o}}(2002)}]{2002AN....323..424L}
{Lanza}, A.~F., \& {Rodon{\`o}}, M. 2002, Astronomische Nachrichten, 323, 424, \dodoi{10.1002/1521-3994(200208)323:3/4<424::AID-ASNA424>3.0.CO;2-1}

\bibitem[{{Li} {et~al.}(2021){Li}, {Xia}, {Kim}, {Hu}, {Guo}, {Jeong}, {Chen}, \& {Gao}}]{2021ApJ...922..122L}
{Li}, K., {Xia}, Q.-Q., {Kim}, C.-H., {et~al.} 2021, \apj, 922, 122, \dodoi{10.3847/1538-4357/ac242f}

\bibitem[{{Liao} \& {Qian}(2010)}]{Liao2010}
{Liao}, W.~P., \& {Qian}, S.~B. 2010, \mnras, 405, 1930, \dodoi{10.1111/j.1365-2966.2010.16584.x}

\bibitem[{{Lohr} {et~al.}(2012){Lohr}, {Norton}, {Kolb}, {Anderson}, {Faedi}, \& {West}}]{Lohr2012}
{Lohr}, M.~E., {Norton}, A.~J., {Kolb}, U.~C., {et~al.} 2012, \aap, 542, A124, \dodoi{10.1051/0004-6361/201219158}

\bibitem[{{Lucy}(1973)}]{Lucy1973}
{Lucy}, L.~B. 1973, \apss, 22, 381, \dodoi{10.1007/BF00647433}

\bibitem[{{Lucy}(1976)}]{Lucy1976}
---. 1976, \apj, 205, 208, \dodoi{10.1086/154265}

\bibitem[{{Maceroni} {et~al.}(1984){Maceroni}, {Milano}, \& {Russo}}]{1984AAS..58..405}
{Maceroni}, C., {Milano}, L., \& {Russo}, G. 1984, \aaps, 58, 405

\bibitem[{{Maceroni} \& {van't Veer}(1996)}]{Maceroni1996}
{Maceroni}, C., \& {van't Veer}, F. 1996, \aap, 311, 523

\bibitem[{{Masci} {et~al.}(2019){Masci}, {Laher}, {Rusholme}, {Shupe}, {Groom}, {Surace}, {Jackson}, {Monkewitz}, {Beck}, {Flynn}, {Terek}, {Landry}, {Hacopians}, {Desai}, {Howell}, {Brooke}, {Imel}, {Wachter}, {Ye}, {Lin}, {Cenko}, {Cunningham}, {Rebbapragada}, {Bue}, {Miller}, {Mahabal}, {Bellm}, {Patterson}, {Juri{\'c}}, {Golkhou}, {Ofek}, {Walters}, {Graham}, {Kasliwal}, {Dekany}, {Kupfer}, {Burdge}, {Cannella}, {Barlow}, {Van Sistine}, {Giomi}, {Fremling}, {Blagorodnova}, {Levitan}, {Riddle}, {Smith}, {Helou}, {Prince}, \& {Kulkarni}}]{2019PASP..131a8003M}
{Masci}, F.~J., {Laher}, R.~R., {Rusholme}, B., {et~al.} 2019, \pasp, 131, 018003, \dodoi{10.1088/1538-3873/aae8ac}

\bibitem[{{Mateo} \& {Rucinski}(2017)}]{2017AJ....154..125M}
{Mateo}, N.~M., \& {Rucinski}, S.~M. 2017, \aj, 154, 125, \dodoi{10.3847/1538-3881/aa8453}

\bibitem[{{McLean} \& {Hilditch}(1983)}]{1983MNRAS.203....1M}
{McLean}, B.~J., \& {Hilditch}, R.~W. 1983, \mnras, 203, 1, \dodoi{10.1093/mnras/203.1.1}

\bibitem[{{Mochnacki}(1981)}]{Mochnacki1981}
{Mochnacki}, S.~W. 1981, \apj, 245, 650, \dodoi{10.1086/158841}

\bibitem[{{Mochnacki} \& {Doughty}(1972)}]{1972MNRAS.156..243M}
{Mochnacki}, S.~W., \& {Doughty}, N.~A. 1972, \mnras, 156, 243, \dodoi{10.1093/mnras/156.2.243}

\bibitem[{{Mochnacki} \& {Whelan}(1973)}]{Mochnacki1973}
{Mochnacki}, S.~W., \& {Whelan}, J.~A.~J. 1973, \aap, 25, 249

\bibitem[{{Mullan}(1975)}]{Mullan1975}
{Mullan}, D.~J. 1975, \apj, 198, 563, \dodoi{10.1086/153635}

\bibitem[{{Nelson}(2010)}]{2010IBVS..5951..1}
{Nelson}, R.~H. 2010, Information Bulletin on Variable Stars, 5951, 1

\bibitem[{{Nelson}(2015)}]{2015IBVS.6142....1N}
---. 2015, Information Bulletin on Variable Stars, 6142, 1

\bibitem[{{Norton} {et~al.}(2011){Norton}, {Payne}, {Evans}, {West}, {Wheatley}, {Anderson}, {Barros}, {Butters}, {Collier Cameron}, {Christian}, {Enoch}, {Faedi}, {Haswell}, {Hellier}, {Holmes}, {Horne}, {Kane}, {Lister}, {Maxted}, {Parley}, {Pollacco}, {Simpson}, {Skillen}, {Smalley}, {Southworth}, \& {Street}}]{Norton2011}
{Norton}, A.~J., {Payne}, S.~G., {Evans}, T., {et~al.} 2011, \aap, 528, A90, \dodoi{10.1051/0004-6361/201116448}

\bibitem[{{O'Connell}(1951)}]{Connell1951}
{O'Connell}, D.~J.~K. 1951, \mnras, 111, 642, \dodoi{10.1093/mnras/111.6.642}

\bibitem[{{Pazhouhesh} \& {Edalati}(2003)}]{2003ApSS.288..259P}
{Pazhouhesh}, R., \& {Edalati}, M.~T. 2003, \apss, 288, 259, \dodoi{10.1023/B:ASTR.0000006042.14745.74}

\bibitem[{{Peng} {et~al.}(2016){Peng}, {Luo}, {Zhang}, {Deng}, {Wang}, {Tian}, {Yan}, {Pan}, {Fang}, {Feng}, {Tang}, {Liu}, {Sun}, \& {Zhou}}]{2016RAA..16..157}
{Peng}, Y.-J., {Luo}, Z.-Q., {Zhang}, X.-B., {et~al.} 2016, Research in Astronomy and Astrophysics, 16, 157, \dodoi{10.1088/1674-4527/16/10/157}

\bibitem[{{Pilecki} \& {Stepien}(2012)}]{2012IBVS.6012....1P}
{Pilecki}, B., \& {Stepien}, K. 2012, Information Bulletin on Variable Stars, 6012, 1

\bibitem[{{Qian} \& {Yang}(2005)}]{Qian+Yang2005}
{Qian}, S., \& {Yang}, Y. 2005, \mnras, 356, 765, \dodoi{10.1111/j.1365-2966.2004.08497.x}

\bibitem[{{Qian} {et~al.}(2005){Qian}, {He}, {Xiang}, {Ding}, \& {Boonrucksar}}]{2005AJ....129.1686Q}
{Qian}, S.-B., {He}, J., {Xiang}, F., {Ding}, X., \& {Boonrucksar}, S. 2005, \aj, 129, 1686, \dodoi{10.1086/427852}

\bibitem[{{Qian} \& {He}(2005)}]{2005PASJ...57..977Q}
{Qian}, S.-B., \& {He}, J.-J. 2005, \pasj, 57, 977, \dodoi{10.1093/pasj/57.6.977}

\bibitem[{{Qian} {et~al.}(2006){Qian}, {Liu}, {Soonthornthum}, {Zhu}, \& {He}}]{2006AJ....131.3028Q}
{Qian}, S.~B., {Liu}, L., {Soonthornthum}, B., {Zhu}, L.~Y., \& {He}, J.~J. 2006, \aj, 131, 3028, \dodoi{10.1086/503561}

\bibitem[{{Qian} {et~al.}(2020){Qian}, {Zhu}, {Liu}, {Zhang}, {Shi}, {He}, \& {Zhang}}]{Qian2020}
{Qian}, S.-B., {Zhu}, L.-Y., {Liu}, L., {et~al.} 2020, Research in Astronomy and Astrophysics, 20, 163, \dodoi{10.1088/1674-4527/20/10/163}

\bibitem[{{Qian} {et~al.}(2015){Qian}, {Zhang}, {Soonthornthum}, {He}, {Rattanasoon}, {Aukkaravittayapun}, {Liu}, {Zhu}, {Zhao}, {Zhou}, \& {Thawicharat}}]{Qian+2015b}
{Qian}, S.~B., {Zhang}, B., {Soonthornthum}, B., {et~al.} 2015, \aj, 150, 117, \dodoi{10.1088/0004-6256/150/4/117}

\bibitem[{{Ricker} {et~al.}(2015){Ricker}, {Winn}, {Vanderspek}, {Latham}, {Bakos}, {Bean}, {Berta-Thompson}, {Brown}, {Buchhave}, {Butler}, {Butler}, {Chaplin}, {Charbonneau}, {Christensen-Dalsgaard}, {Clampin}, {Deming}, {Doty}, {De Lee}, {Dressing}, {Dunham}, {Endl}, {Fressin}, {Ge}, {Henning}, {Holman}, {Howard}, {Ida}, {Jenkins}, {Jernigan}, {Johnson}, {Kaltenegger}, {Kawai}, {Kjeldsen}, {Laughlin}, {Levine}, {Lin}, {Lissauer}, {MacQueen}, {Marcy}, {McCullough}, {Morton}, {Narita}, {Paegert}, {Palle}, {Pepe}, {Pepper}, {Quirrenbach}, {Rinehart}, {Sasselov}, {Sato}, {Seager}, {Sozzetti}, {Stassun}, {Sullivan}, {Szentgyorgyi}, {Torres}, {Udry}, \& {Villasenor}}]{2015JATIS...1a4003R}
{Ricker}, G.~R., {Winn}, J.~N., {Vanderspek}, R., {et~al.} 2015, Journal of Astronomical Telescopes, Instruments, and Systems, 1, 014003, \dodoi{10.1117/1.JATIS.1.1.014003}

\bibitem[{{Rovithis-Livaniou} {et~al.}(2000){Rovithis-Livaniou}, {Kranidiotis}, {Rovithis}, \& {Athanassiades}}]{2000A&A...354..904R}
{Rovithis-Livaniou}, H., {Kranidiotis}, A.~N., {Rovithis}, P., \& {Athanassiades}, G. 2000, \aap, 354, 904

\bibitem[{{Ruci{\'n}ski}(1969)}]{Rucinski1969}
{Ruci{\'n}ski}, S.~M. 1969, \actaa, 19, 245

\bibitem[{{Rucinski}(1974)}]{Rucinski1974}
{Rucinski}, S.~M. 1974, \actaa, 24, 119

\bibitem[{{Rucinski}(1994)}]{1994PASP..106..462R}
---. 1994, \pasp, 106, 462, \dodoi{10.1086/133401}

\bibitem[{{Rucinski} \& {Duerbeck}(1997)}]{1997PASP..109.1340R}
{Rucinski}, S.~M., \& {Duerbeck}, H.~W. 1997, \pasp, 109, 1340, \dodoi{10.1086/134014}

\bibitem[{{Sandquist} \& {Shetrone}(2003)}]{2003AJ..126..2954}
{Sandquist}, E.~L., \& {Shetrone}, M.~D. 2003, \aj, 126, 2954, \dodoi{10.1086/379175}

\bibitem[{{Schieven} {et~al.}(1983){Schieven}, {Morton}, {McLean}, \& {Hughes}}]{1983AAS..52..463}
{Schieven}, G., {Morton}, J.~C., {McLean}, B.~J., \& {Hughes}, V.~A. 1983, \aaps, 52, 463

\bibitem[{{Shappee} {et~al.}(2014){Shappee}, {Prieto}, {Grupe}, {Kochanek}, {Stanek}, {De Rosa}, {Mathur}, {Zu}, {Peterson}, {Pogge}, {Komossa}, {Im}, {Jencson}, {Holoien}, {Basu}, {Beacom}, {Szczygie{\l}}, {Brimacombe}, {Adams}, {Campillay}, {Choi}, {Contreras}, {Dietrich}, {Dubberley}, {Elphick}, {Foale}, {Giustini}, {Gonzalez}, {Hawkins}, {Howell}, {Hsiao}, {Koss}, {Leighly}, {Morrell}, {Mudd}, {Mullins}, {Nugent}, {Parrent}, {Phillips}, {Pojmanski}, {Rosing}, {Ross}, {Sand}, {Terndrup}, {Valenti}, {Walker}, \& {Yoon}}]{2014ApJ...788...48S}
{Shappee}, B.~J., {Prieto}, J.~L., {Grupe}, D., {et~al.} 2014, \apj, 788, 48, \dodoi{10.1088/0004-637X/788/1/48}

\bibitem[{{Stepien}(1980)}]{Stepien1980}
{Stepien}, K. 1980, \actaa, 30, 315

\bibitem[{{Twigg}(1979)}]{1979MNRAS.189..907T}
{Twigg}, L.~W. 1979, \mnras, 189, 907, \dodoi{10.1093/mnras/189.4.907}

\bibitem[{{Van Hamme}(1993)}]{VanHamme1993}
{Van Hamme}, W. 1993, \aj, 106, 2096, \dodoi{10.1086/116788}

\bibitem[{{Wilson}(1978)}]{Wilson1978}
{Wilson}, R.~E. 1978, \apj, 224, 885, \dodoi{10.1086/156438}

\bibitem[{{Wilson}(1990)}]{Wilson1990}
---. 1990, \apj, 356, 613, \dodoi{10.1086/168867}

\bibitem[{{Wilson}(1994)}]{Wilson1994}
---. 1994, \pasp, 106, 921, \dodoi{10.1086/133464}

\bibitem[{{Wilson} \& {Devinney}(1971)}]{Wilson+Devinney1971}
{Wilson}, R.~E., \& {Devinney}, E.~J. 1971, \apj, 166, 605, \dodoi{10.1086/150986}

\bibitem[{{Worthey} \& {Lee}(2011)}]{Worthey+2011}
{Worthey}, G., \& {Lee}, H.-c. 2011, \apjs, 193, 1, \dodoi{10.1088/0067-0049/193/1/1}

\bibitem[{{Xiang} \& {Zhou}(2004{\natexlab{a}})}]{2004NewA....9..273X}
{Xiang}, F.~Y., \& {Zhou}, Y.~C. 2004{\natexlab{a}}, \na, 9, 273, \dodoi{10.1016/j.newast.2003.12.001}

\bibitem[{{Xiang} \& {Zhou}(2004{\natexlab{b}})}]{2004NA..9..273}
---. 2004{\natexlab{b}}, \na, 9, 273, \dodoi{10.1016/j.newast.2003.12.001}

\bibitem[{{Yang} \& {Liu}(2000)}]{yang2000}
{Yang}, Y., \& {Liu}, Q. 2000, \aaps, 144, 457, \dodoi{10.1051/aas:2000219}

\bibitem[{{Yildiz} \& {Do{\u{g}}an}(2013)}]{Yildiz2013}
{Yildiz}, M., \& {Do{\u{g}}an}, T. 2013, \mnras, 430, 2029, \dodoi{10.1093/mnras/stt028}

\bibitem[{{Zhang} {et~al.}(2017){Zhang}, {Yu}, {Xiang}, \& {Hu}}]{2017RAA....17..128Z}
{Zhang}, X.-D., {Yu}, Y.-X., {Xiang}, F.-Y., \& {Hu}, K. 2017, Research in Astronomy and Astrophysics, 17, 128, \dodoi{10.1088/1674-4527/17/12/128}

\bibitem[{{Zhu} {et~al.}(2019){Zhu}, {Wang}, {Tian}, {Li}, \& {Gao}}]{2019MNRAS.489.2677Z}
{Zhu}, L.~Y., {Wang}, Z.~H., {Tian}, X.~M., {Li}, L.~J., \& {Gao}, X. 2019, \mnras, 489, 2677, \dodoi{10.1093/mnras/stz2294}

\end{thebibliography}
\bibliographystyle{aasjournal}

\end{document}